\documentclass[a4paper,fleqn,usenatbib]{mnras}
\pdfoutput=1

\usepackage{times}              
\usepackage[fleqn]{amsmath}     
\usepackage{amssymb}            
\usepackage{graphicx}
\usepackage{booktabs}

\usepackage[usenames, dvipsnames]{color}
\usepackage{rotating}

\usepackage{multirow}

\usepackage[normalem]{ulem}

\def\araa{ARA\&A}%
\def\apj{ApJ}%
\def\apjl{ApJ}%
\def\apjs{ApJS}%
\def\ao{Appl.\ Opt.}%
\def\aap{A\&A}%
\def\mnras{MNRAS}%
\def\prd{Phys.\ Rev.\ D}%
\def\ltsima{$\; \buildrel < \over \sim \;$}
\def\simlt{\lower.5ex\hbox{\ltsima}}   
\def\gtsima{$\; \buildrel > \over \sim \;$}
\def\simgt{\lower.5ex\hbox{\gtsima}}

\def\rhodm{\rho_\mathrm{dm}}

\newcommand{\gevcm}{{\rm GeV\, cm}^{-3}}
\newcommand{\Msunpc}{{\rm M}_\odot \, {\rm pc}^{-3}}
\newcommand{\MsunpcSq}{{\rm M}_\odot \, {\rm pc}^{-2}}
\newcommand{\sd}{\mathit{SD}}
\newcommand{\kmspc}{{\rm km}^2\, {\rm s}^{-2}\, {\rm pc}^{-1}}

\def \MultiNest{{{\sc MultiNest\,}}}

\def \eg{{\textit{e.g.~}}}
\def \ie{{i.e.~}}

\title{The Local Dark Matter Density from SDSS-SEGUE G-dwarfs}
\author[Sivertsson et al.]{ S. Sivertsson$^{1}$, H. Silverwood$^{2,3}$, J. I. Read$^{4}$, G. Bertone$^{3}$, P. Steger$^{5}$\\
$^1$ The Oskar Klein Centre for Cosmoparticle Physics, Department of Physics, Stockholm University, AlbaNova, SE-106 91 Stockholm, Sweden\\
$^2$ Institut de Ci\`encies del Cosmos (ICCUB), Universitat de Barcelona (IEEC-UB), Mart\`i Franqu\`es 1, E08028 Barcelona, Spain\\
$^3$ GRAPPA, University of Amsterdam, Science Park 904, 1098 XH Amsterdam, The Netherlands\\
$^4$ Department of Physics, University of Surrey, Guildford, GU2 7XH, Surrey, UK\\
$^5$ Institute for Astronomy, Department of Physics, ETH Z\"urich, Wolfgang-Pauli-Strasse 27, CH-8093 Z\"urich, Switzerland\\
}

\date{Accepted XXX. Received YYY; in original form ZZZ}
\pubyear{2017}

\begin{document}
\label{firstpage}
\pagerange{\pageref{firstpage}--\pageref{lastpage}}
\maketitle

\begin{abstract}
We derive the local dark matter density by applying the integrated Jeans equation method from \citet{Silverwood:2015} to SDSS-SEGUE G-dwarf data processed and presented by  \citet{Budenbender}. We use the \MultiNest Bayesian nested sampling software to fit a model for the baryon distribution, dark matter and tracer stars, including a model for the `tilt term' that couples the vertical and radial motions, to the data. The $\alpha$-young population from \cite{Budenbender} yields the most reliable result of $\rhodm = 0.46^{+0.07}_{-0.09}\, \gevcm = 0.012^{+0.001}_{-0.002} \Msunpc$. Our analyses yield inconsistent results for the $\alpha$-young and $\alpha$-old data, pointing to problems in the tilt term and its modelling, the data itself, the assumption of a flat rotation curve, or the effects of disequilibria.        
\end{abstract}
 

\section{Introduction}\label{sec:introduction}
The local dark matter (DM) density is a crucial ingredient in the interpretation of results from in-laboratory searches for cosmological DM, be it WIMPs \citep[\eg][]{Baudis:2013eba, 2016JPhG...43a3001M} , sterile neutrinos \citep{2016PhRvD..94i5010C}, or axions \citep[\eg][]{2015ARNPS..65..485G}. In each of these cases the measured signal is degenerately dependent on the local DM density and the DM particle's coupling to the Standard Model. The latter is of immense interest to the study of the beyond-the-Standard Model theory underlying DM, and incorrect conclusions on these models could result from systematic uncertainties on the local DM density. The local DM density is also of interest for measuring the local halo shape, testing galaxy formation theory, and probing alternative gravity models. For a review on the topic of the local DM density see \cite{Read:2014}.

Here we apply the method for determining the local DM density presented in \cite{Silverwood:2015} to G-dwarf data originating from the Sloan Digital Sky Survey and presented in \citet[][hereafter referred to as B15]{Budenbender}. In Section \ref{sec:data} we give further details on this data set, then in Section \ref{sec:method} we present our method and the updates it has undergone since its initial publication in \cite{Silverwood:2015}. The kinematics of stars is dictated by the gravitational potential generated by the sum of dark and baryonic matter, and so to extract the local DM density we must have a good description of the baryonic matter distribution. Thus in Section \ref{sec:baryon_modelling} we present the details of the model we use for the baryonic mass distribution. As in previous work we make the approximation that the DM density is constant with height above the mid-plane\footnote{Up to a height of $z \sim 3$ kpc the midplane value is within 10\% of the value given by a spherical NFW halo with a scale radius of 20 kpc. The impact of this assumption is discussed further in Section \ref{sec:integrated_jeans}.}. In Section \ref{sec:results} we present our results, finding the most robust measurement of the local DM density from this data set to be $\rho_{\rm DM} = 0.46^{+0.07}_{-0.09}\, \gevcm = 0.012^{+0.001}_{-0.002} \Msunpc$.

\section{Data}\label{sec:data}
The input data for this analysis are the tracer densities and velocity dispersions for two stellar populations presented in B15, specifically those shown in figs. 3 and A1 (filled squares) of B15. The raw observations used to generate these tracer densities and velocity dispersions come from an SDSS/SEGUE G-dwarf data set originally presented in \cite{2012MNRAS.425.2144L}, supplemented with proper motions from the USNO-B survey. From this data set B15 extracted two populations defined by iron and $\alpha$-element abundances, \ie ${\rm [Fe/H]}$ and $[\alpha/{\rm Fe}]$. These abundances, especially the $[\alpha/{\rm Fe}]$ abundance, were previously thought to be a proxy for stellar age \citep{2016ApJ...823..114N}, and hence B15 referred to these the two populations as $\alpha$-young and $\alpha$-old, a nomenclature we follow. The $\alpha$-old population has metallicity limits of $0.3 < [\alpha/{\rm Fe}] $ and $-1.2 < {\rm [Fe/H]} < -0.3$ , while the $\alpha$-young population has limits $ [\alpha/{\rm Fe}] < 0.2$ and $-0.5 < {\rm [Fe/H]} $, see fig. 3 in B15. 

We also apply the sign correction to the $\overline{v_R v_z}$ velocity dispersion data from B15 as discussed in Section \ref{tilt_term_section} and in \cite{Silverwood:2015}. B15 also makes the assumption, as we also do, that the Milky Way is symmetric about $z=0$, and so combine data from above and below the plane. We will discuss the validity of this assumption further in Section \ref{sec:disequilibria}.

\section{Method} \label{sec:method}
The broad flow of the method remains the same as that of \cite{Silverwood:2015}. We first make parametrized models of the baryon density distribution $\rho_{\rm baryon}(z)$, the constant DM density $\rho_{\rm DM}(z) = \rho_{\rm DM}$, the tracer star density $\nu(z)$, and the tilt term, which describes the coupling of vertical and radial motions for the tracer populations and incorporates the radial-vertical cross term of the velocity dispersion tensor $\sigma_{Rz}$. Using the integrated $z$-direction Jeans equation described below (equation \ref{key-eq}), we can then derive the vertical velocity dispersion $\sigma_z(z)$ for a given set of parameter values. Using Bayesian nested sampling as implemented by \MultiNest \citep{Feroz:2007kg, Feroz:2008xx, Feroz:2013hea,2014A&A...564A.125B} we scan through the parameter space, with a likelihood function given by
\begin{equation}
\mathcal{L}(\theta) = \exp{\left(- \frac{\chi^2_\nu + \chi^2_{\sigma_{z}} + \chi^2_{\sigma_{Rz}}}{2} \right)},
\label{eqn:chi2_likelihood}
\end{equation}
where
\begin{align}
\chi^2_\nu &= \sum_j \frac{(\nu_{{\rm data},j} - \nu_{{\rm model},j})^2}{\sd_{\nu,j}^2}, \label{eqn:chi2_nu}\\
\chi^2_{\sigma_z} &= \sum_j \frac{(\sigma_{z,{\rm data},j} -\sigma_{z,{\rm model},j} )^2}{\sd_{{\sigma_z},j}^2}, \label{eqn:chi2_sigz}\\
\chi^2_{\sigma_{Rz}} &=\sum_j \frac{(\sigma_{Rz,{\rm data},j} -\sigma_{Rz,{\rm model},j} )^2}{\sd_{{\sigma_{Rz}},j}^2}. \label{eqn:chi2_sigRz}
\end{align}
The $j$ variable iterates over the data points, and $\sd$ is the standard deviation of each data point (\eg $1 \sigma$ errors). From this scan we derive a marginalized posterior on the local DM density $\rho_{\rm DM}$.  

\subsection{The integrated Jeans Equation}\label{sec:integrated_jeans}
The key equation of this method is derived from the Jeans Equations. Assuming dynamical equilibrium (steady state) and axisymmetry, the $z$-Jeans equation in cylindrical coordinates becomes \citep{Binney2008, Silverwood:2015}:
\begin{equation}
\frac{1}{\nu}\frac{\partial}{\partial z}(\nu\sigma_z^2) + \underbrace{\frac{1}{R\nu}\frac{\partial}{\partial R}(R\nu\sigma_{Rz})}_{\mathrm{tilt \ term: \ }\mathcal{T}}  = -\frac{\partial\Phi}{\partial z}, \label{Jeans-eq}
\end{equation}
where $z$ is the perpendicular distance to the galactic plane, $\nu$ is the number density of tracer stars, $\Phi$ is the gravitational potential and $\sigma_z$ is the tracer stars' velocity dispersion in the $z$-direction. In steady state the galactic disc can not oscillate, implying that $\nu$, $\sigma_z$, and $\Phi$ (and thus the underlying mass distribution also) must all be symmetric around $z=0$. Similarly steady state implies that $\bar v_z=0$ and hence we use $\sigma_{Rz}(z)=\overline{v_z v_R}$. The assumptions of steady state and axisymmetry are linked; any deviation from axisymmetry, such as spiral arms, will always imply that the system is evolving with time.  

The Poisson equation in cylindrical coordinates is:
\begin{equation}
\frac{\partial^2 \Phi}{\partial z^2} + \underbrace{\frac{1}{R}\frac{\partial}{\partial R}\left(R\frac{\partial\Phi}{\partial R}\right)}_{\mathrm{rotation \ curve \ term: \ }\mathcal{R}} = 4\pi G\rho,\label{poisson_eq}
\end{equation}
where $\rho(z)$ is the total local mass density. As for equation (\ref{Jeans-eq}), the $\varphi$ derivative term has been excluded as we still assume steady state and hence axisymmetry.

Equation~(\ref{poisson_eq}) includes the circular velocity $V_c$ through $V_c^2 = R\partial\Phi /\partial R$. For a flat rotation curve the circular velocity is independent of $R$ and hence the rotation curve term $\mathcal{R}$ vanishes. If the rotation curve is not completely flat, $\mathcal{R}$ will appear as a small shift of the recovered density, see \cite{Garbari:2012} and \cite{Silverwood:2015}. We will for now neglect $\mathcal{R}$ in the analysis and return to it in Section~\ref{rotation_curve_term_section}.

As in \cite{Silverwood:2015} we then arrive at the key equation by integrating Eq. \ref{Jeans-eq} with respect to $z$:
\begin{equation}
\nu(z)\sigma_z^2(z) =  \nu(z_0)\sigma_z^2(z_0) - \int_{z_0}^z \nu(z')[2\pi G\Sigma(z')+\mathcal{T}(z')]\mathrm{d}z'. \label{key-eq}
\end{equation}
We assume the solar neighborhood to be symmetric above and below the disc plane, and hence the surface density is given by $\Sigma(z) = 2\int_0^z\rho(z)\mathrm{d}z$, where $\rho(z)$ is the local density. 
In \cite{Silverwood:2015} we used a normalization parameter $C\equiv \sigma_z^2(z_0)\nu(z_0)$, where $C$ was treated as a free parameter that the code marginalized over given some prior range. This worked well for good data which fit the model well. However, to treat $C$ as a free parameter also for data which have some tension in it gives artificial freedom to find better-fitting unphysical solutions, because in reality the $C$ constant is not a free parameter.

As seen in fig. 3 of B15, the tracer density data for both these populations are well fit by one exponential function: $\nu(z) \propto \exp(-z/h)$ each. We will hence assume exponential shapes of the tracer densities for the different populations; note that the normalization of $\nu(z)$ cancels in equation (\ref{key-eq}), and hence for each tracer population we are only interested in the scale height $h$.

Assuming $\nu(z)\propto{\rm e}^{-z/h}$, one sees that $\nu(z_0)$ falls quickly towards 0 for $z_0\to\infty$, resulting in $\nu(z_0)\sigma_z^2(z_0)\to 0$ for $z_0\to\infty$. Hence in the limit $z_0\to\infty$, equation (\ref{key-eq}) becomes:
\begin{equation}
\sigma_z^2(z) = \mathrm{e}^{z/h}\int_z^\infty \mathrm{e}^{-z'/h}[2\pi G\Sigma(z')+\mathcal{T}(z')] \mathrm{d}z'. \label{analytic0}
\end{equation}
The integrand of equation (\ref{analytic0}) falls quickly enough with $z$ for the integral to be finite\footnote{This is the case as long as the tilt term is well behaved like our model and does not increase fast enough to overwhelm the $\exp(-z/h)$ term.} even though the integration runs to infinity. 

That the integral in equation (\ref{analytic0}) runs all the way to infinity implies that $\sigma_z^2(z)$ depends on the tracer and mass density distribution all the way out to infinitely large $z$. However, as $\nu(z)\propto{\rm e}^{-z/h}$ falls quickly with $z$, the contribution to $\sigma_z^2(z)$ from the integral in equation (\ref{analytic0}) over large $z$ becomes very small\footnote{The total density decreases with $z$ and hence the surface density cannot increase faster than $\Sigma(z)\propto z$, which would result in a primitive function  $F \propto \int {e}^{-z}\Sigma(z)\mathrm{d}z  \propto -(z+1)e^{-z}\to 0$ as $z\to\infty$. Hence for all realistic surface density profiles there always exists a primitive function $F(z)$ such that $F(z)\to 0$ as $z\to\infty$.}. For example, in this paper the dark matter density is assumed to be constant for all $z$, rendering an infinite dark matter surface density as $z\to\infty$. To quantify the effect on $\sigma_z^2$ of this over-prediction of the dark matter density at large $z$, one can compare with the resultant $\sigma_z^2$ when making the more realistic assumption that the galactic dark matter density is proportional to $1/R$, where $R$ is the distance to the Galactic centre. For the tracer data used in this paper the difference in $\sigma_z^2$ between these two dark matter density assumptions is maximized for the largest $z$ data bin and the $\alpha$-old data. However, this maximal difference in the dark matter contribution to $\sigma_z^2$ is still only 2.4\%, and the relative difference is of course reduced further when also taking into account the baryonic contribution. The smallness of the large $z$ contributions to $\sigma_z^2$ is further discussed in Sections \ref{sigma_z_without_tilt_section} and \ref{tilt_contribution_to_sigma_z_section}. 

The integral in equation (\ref{analytic0}) can be divided into one integral over $\Sigma(z)$ and one over $\mathcal{T}(z)$; we will first look at $\sigma_z^2$ when neglecting the tilt term and then add the tilt term contribution to $\sigma_z^2$.

\subsection{Analysis of $\sigma_z^2$ without the tilt term} \label{sigma_z_without_tilt_section}
When neglecting tilt, equation (\ref{analytic0}) becomes:
\begin{equation}
\sigma_z^2(z) = \mathrm{e}^{z/h}2\pi G\int_z^\infty \mathrm{e}^{-z'/h}\Sigma(z') \mathrm{d} z'. \label{sigz2_no_tilt_eq}
\end{equation}
$\Sigma(z)$ is the total surface density which can further be divided into the dark matter contribution $\Sigma_{\rm{dm}}=2\rho_{\rm{dm}}z$, and the baryonic contribution $\Sigma_{\rm{baryon}}$. Given data in the range $[z_-, z_+]$, with $z_-, z_+ > 0$, the baryonic contribution can be divided into components which contribute significantly to the total density above $z_-$, labeled $\Sigma_{b1}$, and those which do not, \eg are concentrated close to the disc plane below $z_-$, labeled $\Sigma_{b0}$. For this latter component, above $z_-$ the density is close enough to zero for the surface density to be considered constant. Hence, for $z>z_-$, the total surface density is $\Sigma(z)=\Sigma_{b0} + \Sigma_{b1}(z) + 2\rho_{\mathrm{dm}}z$, and equation (\ref{sigz2_no_tilt_eq}) then becomes:
\begin{equation}
\frac{\sigma_z^2(z)}{2\pi Gh} = \Sigma_{b0} + \underbrace{2\rho_{\mathrm{dm}}(h+z)}_{=\Sigma_{\mathrm{dm}}(z) + \Sigma_{\mathrm{dm}}(h)} + \frac{\mathrm{e}^{z/h}}{h}\int_z^\infty\Sigma_{b1}(z')\mathrm{e}^{-z'/h}\mathrm{d}z. \label{sigz2_not_tilt2_eq}
\end{equation}
Note that the baryonic density shape inside $z$ does not enter in the calculation of $\sigma_z(z)$; hence there is no need to model the shapes of the baryonic components inside the $z$ region where we have tracer data.

Also note the term $2\rho_{\mathrm{dm}}(h+z)$ of equation (\ref{sigz2_not_tilt2_eq}) can be expressed as $\Sigma_{\mathrm{dm}}(z) + \Sigma_{\mathrm{dm}}(h)$. Despite having a constant DM density to an infinite height the contribution of this component to $\sigma_z^2$ is finite. The term $\Sigma_{\mathrm{dm}}(h)$ is a manifestation of the tracer density decaying exponentially with a scale height of $h$. Thus the impact of the assumption of constant DM density is controlled by the scale height of the tracer population, with hotter populations reaching higher above the disc being affected more. The contribution of

\subsection{Tilt term} \label{tilt_term_section}
At the midplane ($z=0$) a potential symmetric in $z$ is separable up to second order \citep{Binney2008}, with the radial and vertical motions decoupling. Thus at $z=0$ the tilt term vanishes, though it may increase rapidly as $z$ increases.   

Intuitively, the motion of the tracer stars can be viewed as a combination of oscillation around the disc, and orbital motion around the galactic centre; the tilt term can then be seen as a consequence of this orbital motion around the Galactic centre. A star in an elliptic orbit around the Galactic centre will on the part of its orbit where it is moving towards the galactic centre (i.e. has $v_R<0$) typically also move towards the disc plane; hence this star will for $z>0$ have $\overline{v_z v_R}>0$, and $\overline{v_z v_R}<0$ for $z<0$. Similarly, in the part of the orbit where the star moves away from the Galactic centre (i.e. has $v_R>0$) it will typically also move away from the disc plane, again resulting in $\overline{v_z v_R}$ having the same sign as $z$. Hence, as $\sigma_{Rz}=\overline{v_z v_R}$, we arrive at the result that $\sigma_{Rz}$ carries the same sign as $z$. See \cite{Silverwood_thesis} for further motivation on this result on the sign of $\sigma_{Rz}$. 

The $\sigma_{Rz}$ data given in B15 has a sign error, as it is negative for positive $z$, and so as done in \cite{Silverwood:2015} we apply a sign correction. We again assume symmetry above and below the disc plane, and for $z>0$ we model the vertical profile of $\sigma_{Rz}$ using the functional form $\sigma_{Rz}=Az^n$, where $A$ and $n$ are fitted to the $\sigma_{Rz}$ data. For $A\geq 0$ and $n>0$, this model naturally encompasses the requirements that $\sigma_{Rz}\geq 0$ for $z\geq 0$ and $\sigma_{Rz}=0$ for $z=0$. The modelling $\sigma_{Rz}=Az^n$ also fits the $\sigma_{Rz}$ data well, see Fig.~\ref{2pop_fig} and \cite{Silverwood:2015}.

The tilt term includes a radial derivative of $\nu$ and $\sigma_{Rz}$, and so we must also model their radial behaviours. From \cite{Bovy2016} we see that, at the solar location $R=R_\odot$, the tracer densities for the different populations are well fit by an exponential function $\nu(R,z)=\nu(R_\odot,z)\exp(-k_0(R-R_\odot))$. We assume that also $\sigma_{Rz}$ is separable in our local $R$ region and locally parametrize it as $\sigma_{Rz}(R,z)=\sigma_{Rz}(R_\odot,z)\exp(-k_1(R-R_\odot))$, which automatically fulfills the requirement $\partial\sigma_{Rz}/\partial R \to 0$ for $z\to 0$.

The tilt term, evaluated at our location: $R=R_\odot$, then becomes:
\begin{align}
\mathcal{T}(R_\odot,z) &\equiv \frac{1}{R\nu}\frac{\partial}{\partial R}(R\nu\sigma_{Rz}) \nonumber \\
&= \left(\frac{1}{R_\odot} - k\right)\sigma_{Rz}(z) = \underbrace{\left(\frac{1}{R_\odot} - k\right)}_{\equiv K}Az^n, \label{tilt_eq}
\end{align}
where $k\equiv k_0+k_1$, and $R_\odot\simeq 8$~kpc is our distance to the galactic centre.

\subsubsection{Contribution of $\mathcal{T}$ to $\sigma_{z}^2$} \label{tilt_contribution_to_sigma_z_section}
With this we can now calculate the tilt contribution to $\sigma_z^2$, which we will refer to as $\sigma_{z,\mathcal{T}}^2$. From equation (\ref{tilt_eq}) we have  $\mathcal{T}(R_\odot,z)=KAz^n$, and using equation (\ref{key-eq}) with $z_0\to 0$ we then get:
\begin{eqnarray}
e^{-z/h}\sigma_{z,\mathcal{T}}^2(z) &=& \sigma_{z,\mathcal{T}}^2(0) - KA\int_{0}^z t^ne^{-t/h}\mathrm{d}t \nonumber \\
&=& \sigma_{z,\mathcal{T}}^2(0) - KAh^{n+1}\Gamma_{\mathrm{inc}}\left(n+1,\frac{z}{h}\right), \label{sigma_tilt_eq1}
\end{eqnarray}
where $\Gamma_{\mathrm{inc}}$ is the lower incomplete gamma function.

As for equation~(\ref{analytic0}), equation~(\ref{sigma_tilt_eq1}) must hold for all $z$; in the limit $z\to\infty$ we have $e^{-z/h}\to 0$ and $\Gamma_{\mathrm{inc}}(n+1,z/h)\to\Gamma(n+1)$. Requiring $\sigma_{z,\mathcal{T}}^2(z)$ to be finite for all $z$ then gives the relation $\sigma_{z,\mathcal{T}}^2(0) = KAh^{n+1}\Gamma(n+1)$. Putting this back into equation (\ref{sigma_tilt_eq1}) yields the result:
\begin{equation}
\sigma_{z,\mathcal{T}}^2(z) = KAh^{n+1}\left(\Gamma(n+1)-\Gamma_{\mathrm{inc}}\left(n+1,\frac{z}{h}\right)\right)e^{z/h}.\label{sigz2_tilt}
\end{equation}

\begin{figure}
\includegraphics[scale=0.475, trim={1cm 0.2cm 0 1.1cm},clip]{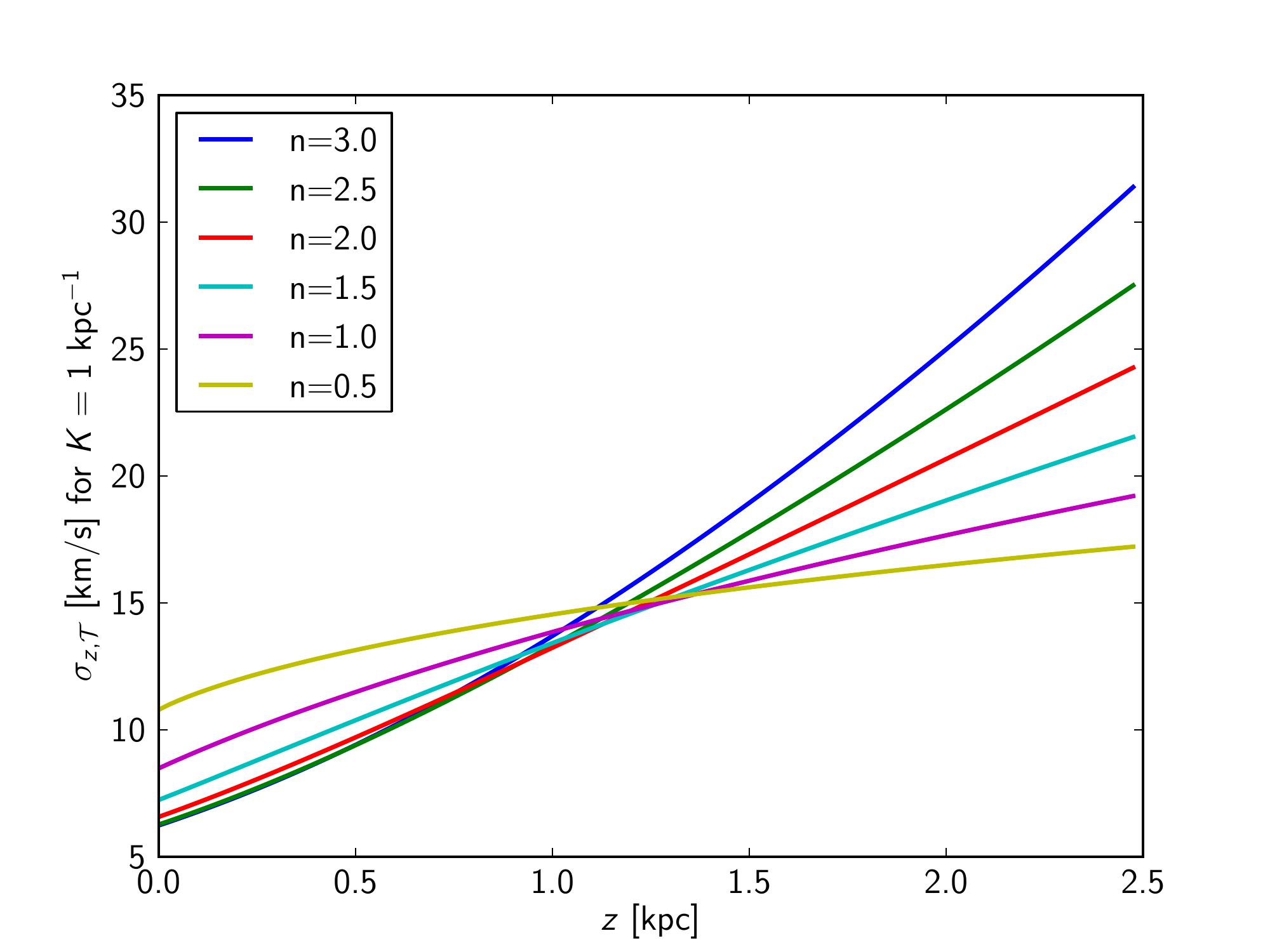}  
\caption{Illustrating the shape of the tilt contribution to $\sigma_z$ by plotting $\sigma_{z,\mathcal{T}}(z)$ for $K=1$~kpc$^{-1}$, for different values of $n$ under the normalization requirement $\sigma_{Rz}(z=2$~kpc$)=400$~(km/s)$^2$, i.e.~$A = 400/2^n$. The normalization of $A$ is chosen so that the $Az^n$ model gives a reasonable fit for all $n$ values to the the $\alpha$-old population's $\sigma_{Rz}$ data, see \eg Fig.~\ref{2pop_fig}. For black and white version of the image: the ordering of the lines on the right hand side of the plot is the same as in the legend. }
\label{sigz_tilt_fig}
\end{figure}

The shape of the tilt contribution to $\sigma_z^2$, based on equation (\ref{sigz2_tilt}), is shown in Fig.~\ref{sigz_tilt_fig}. Note that even though $\mathcal{T}\to 0$ as $z\to 0$, this is not the case for $\sigma_{z,\mathcal{T}}^2(z)$; the $\sigma_{Rz}$ term at higher $z$ affects $\sigma_z^2$ also at lower $z$. 

The data we have on $\sigma_{Rz}$ is quite noisy, as can be seen in \eg Fig.~\ref{2pop_fig}, and hence there is some freedom in the value of $n$ when fitting to the $\sigma_{Rz}$ data. A change in the value of $n$ does however have a large impact on the value of $Az^n$ for large $z$, i.e.~outside the region where we have $\sigma_{Rz}$ data. In deriving equation~(\ref{sigz2_tilt}) we take the limit $z\to\infty$ and hence one might worry that any small modification of $n$ will have a disproportionally large impact on $\sigma_{z,\mathcal{T}}^2(z)$ also in the lower $z$ region where we have tracer data. From Fig.~\ref{sigz_tilt_fig}, showing $\sigma_{z,\mathcal{T}}^2(z)$ for different values on $n$, we however see that this is not the case; for the $z$ region where we have tracer data $\sigma_{z,\mathcal{T}}^2(z)$ has similar normalization for all plotted values of $n$. Hence, as was the case for $\sigma_z^2$ when neglecting the tilt term, the exponential fall off of the tracer density, see \eg equation~(\ref{key-eq}), protects $\sigma_z^2(z)$ from a disproportionally large impact at low $z$ from contributions at high $z$ where we do not have tracer data.

\subsubsection{Prior range on $K\equiv (1/R_\odot-k)$} \label{K_prior}
In equation (\ref{sigz2_tilt}) $A$ and $n$ are determined by the fit to the $\sigma_{Rz}$ data. On the other hand $K$, defined in equation~(\ref{tilt_eq}), is not well measured, and hence we will need to impose a prior on it. 

In this paper we use two stellar tracer populations from B15: the $\alpha$-young and $\alpha$-old populations; the former consists of stars with low [$\alpha/$Fe] and the latter are stars with high [$\alpha/$Fe]. In \cite{Bovy2016} they find that the radial surface density profiles of all the mono-abundance populations with high [$\alpha/$Fe] are well described as falling exponential functions with a scale length of $2.2\pm0.2$~kpc. Hence for the $\alpha$-old stars we use that $k_0=1/2.2\simeq 0.45$~kpc$^{-1}$.

For the low [$\alpha/$Fe] populations in \cite{Bovy2016} it is more complicated as the scale length is also dependent on [Fe/H]. In \cite{Bovy2016} they find that for the populations with low [$\alpha/$Fe] and [Fe/H]$\leq 0$, $\nu(R,z)$ is an increasing function of $R$ at the solar location $R_\odot$, contrary to the case for the high [$\alpha$/Fe] populations. From B15 we see that only a very small fraction of the stars in the $\alpha$-young population have [Fe/H]$>0$, and hence we infer from \cite{Bovy2016} that essentially all sub-populations in our $\alpha$-young population have $\nu(R,z)$ which are increasing functions of $R$. From \cite{Bovy2016} this then implies that the $\alpha$-young population has $-0.27\leq k_0\leq 0$~kpc$^{-1}$.

The scale length of $\sigma_{Rz}(R,z)$, and hence $k_1$, are not very well known. To be somewhat generous in our prior ranges we, for both populations, assume that $\sigma_{Rz}(R)$ has a scale length of more than 1~kpc, i.e.~that $-1\leq k_1\leq 1$~kpc$^{-1}$. This condition on $k_1$ simply assures that $\sigma_{Rz}$ cannot change very quickly with $R$. For example, a scale length of $\sigma_{Rz}$ of $<1$~kpc (i.e.~$k_1>1$~kpc$^{-1}$) implies that $\sigma_{Rz}$ changes by a factor of more than 20 over 3 kpc, seemingly in tension with a smooth behavior of $\sigma_{Rz}$. A further motivation on the validity of the $k_1$ prior range is found in Appendix \ref{k1_prior_appendix}.

For $k=k_0+k_1$ we then arrive at the prior ranges: $-1.3\leq k\leq 1$~kpc$^{-1}$ for the $\alpha$-young population and $-0.5\leq k\leq 1.5$~kpc$^{-1}$ for the $\alpha$-old population. Finally, for the prior range on $K\equiv (1/R_\odot-k)$ we use $R_\odot=8$~kpc. Note that by marginalising over $K$ we also marginalise over uncertainties in the galactocentric distance $R_\odot$.

The tilt term is expected to be more important for the $\alpha$-old population than for the $\alpha$-young population; the latter is more confined to the disc plane and hence less sensitive to the global structure of the galaxy. The prior ranges for the $K$ parameter is not too different between the populations; on the other hand $\sigma_{Rz}(z)$, and hence $A$, is larger for the $\alpha$-old than for the $\alpha$-young population for any given $z$, see \eg Fig.~\ref{2pop_fig}. Furthermore the $\alpha$-old population has a larger scale height, making the larger $z$ part of $\sigma_{Rz}(z)$ more important, and hence the tilt term larger as $\sigma_{Rz}(z)$ increases with $z$, see \eg Fig.~\ref{2pop_fig} and recall that $\sigma_{Rz}(z=0)=0$.

\subsubsection{Prior range on $n$ and $A$} \label{n_prior}
The $A$ and $n$ parameters of equation (\ref{sigz2_tilt}) are fitted to the $\sigma_{Rz}$ data using the functional form $\sigma_{Rz}=Az^n$, as previously described in Section \ref{tilt_term_section}. 

\begin{figure}
\includegraphics[scale=0.475, trim={0.7cm 0.2cm 1.0cm 1.2cm},clip]{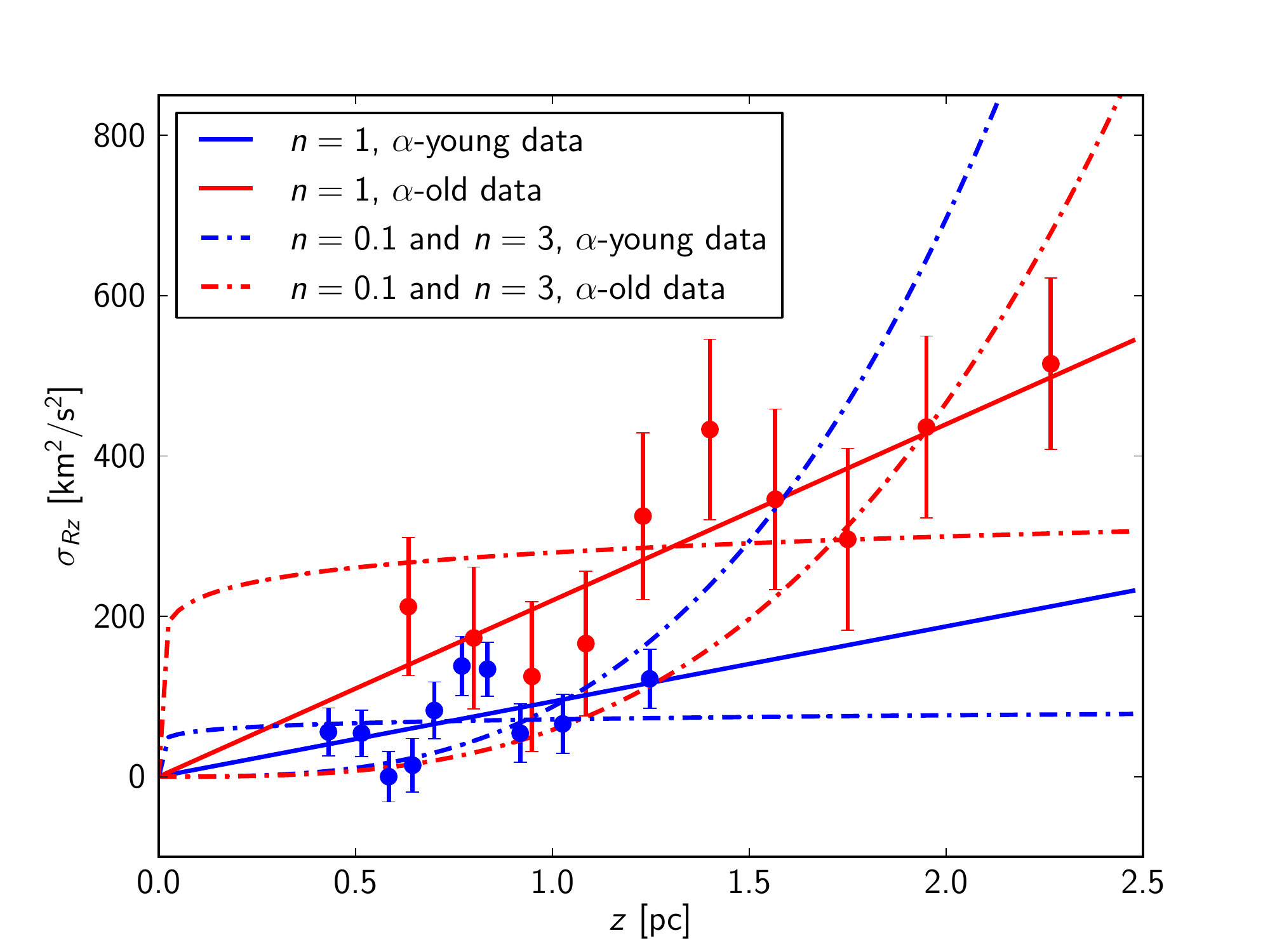}  
\caption{Model fits of the function $Az^n$ to the $\sigma_{Rz}$ to data from B15 (with inverted sign on the data points, see discussion in Section \protect{\ref{sigma_z_without_tilt_section}}). As seen in the plot, the $n=1$ fits (solid lines) fit the data well, while the $n=0.1$ and $n=3$ fits (dot-dashed lines) are not as good fits. This can also be seen by comparing the $\chi^2$-values for the different fits, which for the $\alpha$-young data are: 16.7, 12.9 and 23.0, and for the $\alpha$-old data: 13.4, 4.3 and 23.3, for the $n=0.1$, $n=1$ and $n=3$ fits, respectively.}
\label{A_n_prior_fig}
\end{figure}

The used prior range for $n$ is $0.1<n<3$. As seen in Fig.~\ref{A_n_prior_fig}, the $n=1$ fits fit the data quite well, which is not true for the $n=0.1$ and $n=3$ fits. One exception is the $n=0.1$ fit to the $\alpha$-young data which results in a $\chi^2$-value that is not all that much larger than the $\chi^2$-value for the $n=1$ fit. However, $n=0.1$ seems rather unphysical as its value is more or less constant for all $z$ and then makes a sudden dive to zero close to $z=0$, to fulfill the requirement $\sigma_{Rz}=0$ for $z=0$, as seen in Fig.~\ref{A_n_prior_fig}. Hence we conclude that the used prior range $0.1<n<3$ is generous enough to allow sufficient freedom in fitting the $\sigma_{Rz}$ data.

For the $A$ prior range we use $0\leq A\leq 400\,\kmspc$, where the lower limit comes from the requirement that $\sigma_{Rz}\leq 0$. The largest $A$ value in the fits of Fig.~\ref{A_n_prior_fig} is $A=279.5\,\kmspc$, which is for the fit to the $\alpha$-old data for $n=0.1$, i.e. the lower edge of our prior range on $n$. As seen in Fig.~\ref{A_n_prior_fig}, the $\sigma_{Rz}$ data is well fitted by the $n=1$ lines, these fits result in the $A$ values: $A=93.7$ and $A=219.8\,\kmspc$ for the $\alpha$-young and $\alpha$-old data, respectively, i.e. again well below $A=400$. Hence we conclude that the upper limit of $A\leq 400\,\kmspc$ is sufficiently generous. Finally, as discussed in Sec \ref{young_old_discrepancy_tilt} the $A$ and $n$ prior ranges presented here do not constrain the posterior $A$ and $n$ distributions.

\section{Modelling the baryonic surface density contribution}\label{sec:baryon_modelling}
\begin{figure}
\includegraphics[scale=0.43, trim={0 0 1cm 0},clip]{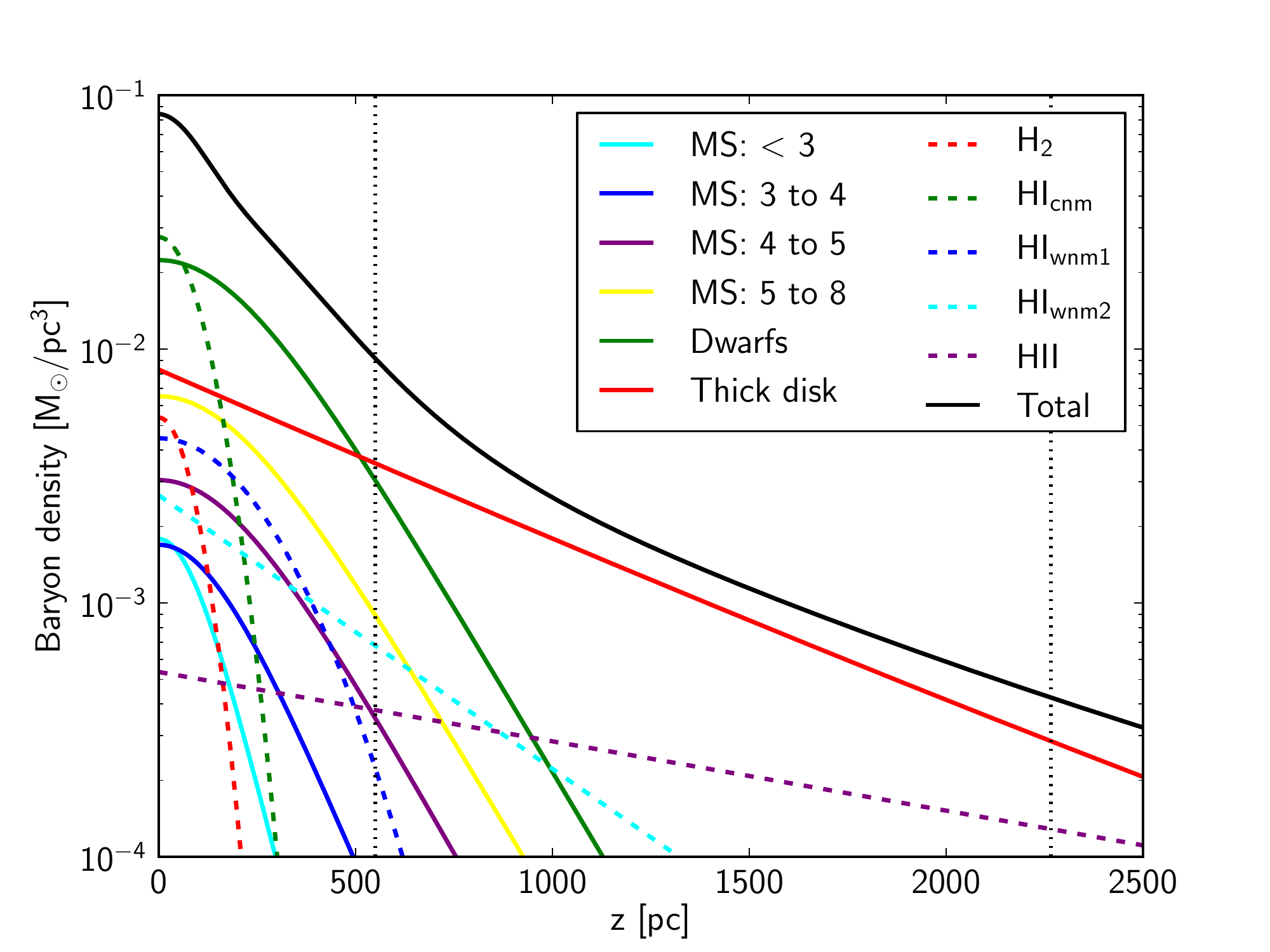} 
\caption{The density profiles of the different baryonic components; dashed lines are gas and solid lines are stellar components, with the solid black line being the total baryonic density (stars + gas). The main sequence stars are labelled with \emph{MS}, followed by their visual magnitude range. The \emph{MS} and \emph{Dwarfs} lines refer only to their thin disc components, while the thick disc components are combined into the red line labelled \emph{Thick disc}. The main sequence stars with $M_V < 4$ are thought to be too young to have a thick disc component, and thus do not contribute to the \emph{Thick disc} line. The dotted vertical lines marks the borders of the region of the data we use. For this plot parameters are mainly the same as in \protect\cite{McKee}, with differences detailed in Section \ref{sec:baryon_modelling}. The \emph{MS: 4 to 5} category include giants, and \emph{Dwarfs} = M dwarfs + brown dwarfs + WD + NS + BH.}
\label{barydens_fig}
\end{figure}

\begin{figure}
\includegraphics[scale=0.43, trim={0 0 1cm 0},clip]{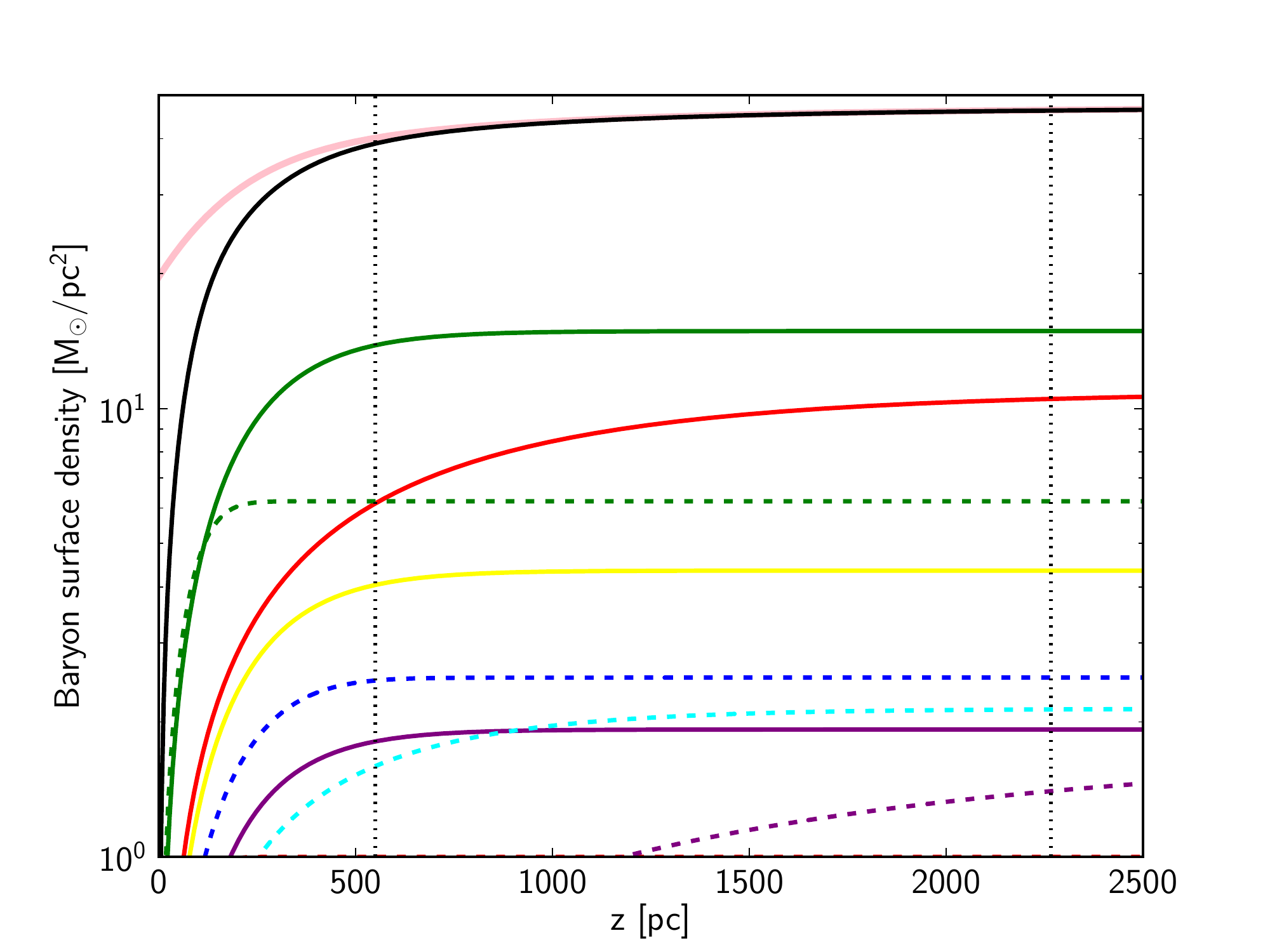} 
\caption{The surface density profiles for the baryonic density profiles shown in Fig.~\ref{barydens_fig}, with the same color coding. The solid pink line is the total surface density for our simplified baryonic model discussed in Section \ref{sec:baryon_modelling} and the black line is the profile of the total surface density, i.e. the sum of all the baryonic components. As seen in the figure the pink line very well mimics the black line within the region of interest between the dotted horizontal lines where we have data. Additionally, there is some freedom in the total baryonic surface density (i.e.~at $z\to\infty$) in the implementation of our baryonic model, which allows the pink line to fit the black line even better.}
\label{barysurfdens_fig}
\end{figure}

A crucial part of determining the local dark matter density is to model the contribution to the surface density from ordinary matter, i.e. baryons. The motions of the tracer stars are dictated by the total potential, and are agnostic as to whether the potential is generated by baryons or DM. Thus we must have a reliable baryon census to detect the extra contribution from DM. Differences in their distributions will help us distinguish the two; baryonic mass is concentrated close to the midplane, while DM extends high above (here we make the simplifying assumption that it is constant in $z$). \cite{1984ApJ...276..169B} and \cite{Garbari:2011} noted that one must go approximately three times the baryon disc scale height above the midplane ($\sim 600$ pc) to break the DM-baryon degeneracy. Also note that the tracer population and the baryon distribution are treated as separate elements in our analysis, \eg the data from B15 is not used to inform the baryon mass distribution model. 

The baryonic budget of the Milky Way can be subdivided into gas and stars. The gas of the Milky Way is dominated by hydrogen, and so the gas component is generally referred to as hydrogen. To account for the presence of heavier elements such as helium~4 the mass of the hydrogen atom is increased during the conversion from number density to mass density. This process is known as the \emph{helium correction}. Following \cite{McKee} we assume the mass in non-hydrogen elements is 40\% of that in hydrogen, thus increasing the effective hydrogen atom mass by a factor of 1.4, \ie $m_{H, {\rm eff}} = 1.4 \times m_H = 2.34\times 10^{-27}$~kg. The gas component is further divided into molecular gas H$_2$, atomic gas HI, and ionized gas HII. 

The stellar component of the Milky Way can be divided into visible stars and stellar remnants. The former consists of main sequence (MS) stars, giants, and M-dwarfs, while the latter is made up of white dwarfs (WDs), brown dwarfs (BDs), neutron stars, and black holes. 

The vertical distribution of each of the gas and stellar components can be described by a functional form fitted to data, with a given number of parameters and associated uncertainties. For instance, M-dwarfs can be described by the sum of a $\mathrm{sech}^2$ and an $\exp$ distribution, which could be thought of as thin and thick disc components respectively. A recent reanalysis of observed baryon data was presented in \cite{McKee}, which we will follow closely. In Figs.~\ref{barydens_fig} and \ref{barysurfdens_fig} we plot respectively the densities and surface densities of the baryonic components, with profiles derived using data from \cite{McKee}, \cite{Z01} and \cite{F06}.

A complete model with a full functional form for each of the components would have several tens of parameters, making it computationally expensive. Fortunately for this analysis we can use a much simpler model due to the fact that the $\sigma_z^2$ value at some height $z = z_i$ as calculated in Eq. \ref{analytic0} only depends on the surface density $\Sigma(z)$ profile for values at or above $z_i$. Thus the the only components that require a full profile are those with significant contributions to the mass density above $z_-$, where our data lies.  The surface densities of all other components can be modeled as a constant surface density component. This can be thought of as placing all the density of these components at $z=0$. 

Our model thus consists of one variable for the total baryon surface density, and then variables describing the surface density and vertical profile of the subcomponents with significant contributions beyond $z_-$. The surface densities of these significant subcomponents is subtracted from the total baryon surface density, and the remainder is taken to be the constant component below $z_-$ made up of the components with insignificant contributions above $z_-$. 

From the density profiles shown in Fig.~\ref{barydens_fig} we can see which components play a significant role above $z_-$. These are the thin disc dwarf component (green solid line covering M dwarfs, white dwarfs, brown dwarfs, neutron stars and black holes); the joint thick disc component, consisting of the thick disc components of MS stars with $M_V > 4$ (including giants) and dwarfs (red solid line); and the HII gas component (purple dashed line). MS stars with $M_V < 4$ are considered too young to have a thick disc component. Thus our baryon model is the sum of these three components, plus a constant surface density component representing the sum of all other components. 

In Fig.~\ref{barysurfdens_fig} we plot the surface densities of many of these baryonic components. The complete unsimplified profile is shown in black and is a summation of all other lines, except the pink line which is our simplified model. While diverging from the complete profile at low $z$, our simplified model converges completely with it by $z=z_-$, illustrating the validity of our simplified model. 

\subsection{Surface density values}
\begin{table*}\label{tab:surfdens_baryons}
\caption{Compilation of baryon surface density values and uncertainties used in this work, and their respective sources. M15 denotes \citet{McKee}, while S17 refers to this work (\eg our own derivations or re-derivations). For the visible stars uncertainty we take the 15\% error quoted by \citet{McKee} in the text as opposed to the 10\% apparently used in Eq. 36 of that work.}
\begin{tabular}{l c c c c }
\hline
Component & $\Sigma^\infty \, \MsunpcSq$ & $\pm$ (\%) && $\pm (\MsunpcSq)$ \\
\hline
Visible stars & 27.0 (M15) & 15\% (M15, text) & $\rightarrow$ & 4.05 \\
White dwarfs & 4.7 (S17) &  17\% (M15) & $\rightarrow$ & 0.80 \\
Brown dwarfs & 1.2 (M15) & 30\% (M15) & $\rightarrow$ & 0.36 \\
Neutron stars & 0.2 (M15) & 30\% (M15) & $\rightarrow$ &0.06 \\
Black holes & 0.1 (M15) & 30\% (M15) & $\rightarrow$ & 0.03 \\
\textbf{Stellar sum} & \textbf{33.2} & & & \textbf{5.30}\\
\\
H2 & 0.95 (S17) & 30\% (M15) & $\rightarrow$ & 0.29 \\
HI & 10.9 (M15) & 20\% (S17) & $\rightarrow$ & 2.18 \\
HII & 1.8 (M15) & 17\% (S17) & $\rightarrow$ & 0.31 \\
\textbf{Gas sum} & \textbf{13.65} & & & \textbf{2.78}\\
\\
\textbf{Total baryon} & \textbf{46.85} & \textbf{13 \%}  & $\leftarrow$ & \textbf{5.98}\\
\hline
\end{tabular}
\end{table*}

For the exact profiles of our baryon components we draw heavily from \cite{McKee}, but include information from several other sources. A summary of the surface density values and uncertainties we take is given in Table~\ref{tab:surfdens_baryons}. For the total surface density we have $\Sigma_{\rm baryon}^\infty = 46.95 \,\MsunpcSq \pm 13\%$, compared to the value from \cite{McKee} of $\Sigma_{\rm baryon}^\infty = 47.1 \,\MsunpcSq \pm 7\%$. The slight difference in the absolute numbers and uncertainties results from different calculations and roundings in the gas components and the white dwarf component, which we will discuss below.  For the total surface mass density we take a flat prior in the range [46.95 - 13\%, 46.95 + 13\%].   
 
\subsubsection{White dwarf surface density}
The local white dwarf density has been measured as $4.9\times 10^{-3}$~WD/pc$^3$ by \cite{Sion2009}, and $6.0\times 10^{-3}$~WD/pc$^3$ by \cite{Reid2005}. \cite{Sion2009} also estimates the local number density of single (i.e. not in binary systems) white dwarfs to be  $3.3\times 10^{-3}$~WD/pc$^3$. There is however a tension between the high fraction of binary systems among MS stars and the low number of binaries of white dwarfs, which are the end product of most MS stars. \cite{Katz14} point out that the number of observed bright WDs with a MS companion is about the same as the number of bright single WDs, while the number of faint WDs with companions is much less than the observed faint single WDs. In \cite{Katz14} they estimate that about 60\% of the WDs are in binary system; using the density of single WDs above from \cite{Sion2009} this gives a WD number density of $8.25\times 10^{-3}$~WD/pc$^3$. Using the \cite{Holberg2008} mean value of the WD mass of 0.665~M$_\odot$, we arrive at a local WD mass density of $\rho_{\rm WD} = 0.0055$~$M_\odot$pc$^{-3}$. \cite{McKee} estimates the number density of white dwarfs by looking at the historical star formation rate and initial mass function (IMF), from which they arrive at a local WD number density of about $8.45\times 10^{-3}$~WD/pc$^{3}$, and thus a local mass density of $\rho_{\rm WD} = 0.0057\, \Msunpc$ when they take the same value for the mean WD mass as we do. To convert to a total surface density we take a scale height of $h = 430$ pc, as listed in table 1 of \cite{McKee}, to arrive at $\Sigma_{\rm WD}^\infty = 2 \rho_{\rm WD} h_{\rm WD} = 4.7\, \MsunpcSq$. \cite{McKee} on the other hand use $h_{\rm WD} = 434$ pc (as they list in the text of their paper), to arrive at $\Sigma_{\rm WD}^\infty = 4.9\, \MsunpcSq$. We take the same percentage uncertainty on our surface density value as \cite{McKee}. 

\subsubsection{Molecular gas H$_2$ surface density}
The slight difference in H2 surface density is due to a recalculation of the \cite{McKee} results that has been rounded to 2 decimal places rather than 1. 

\subsubsection{Atomic gas HI surface density}
For the uncertainty on HI surface density \cite{McKee} assumes 20\% for the Cold Neutral Medium (CNM) component, and for the two Warm Neutral Medium (WNM) components assume 10\% working from the \cite{Heiles1981} stray radiation correction. The CNM has a surface density of $6.21\,\MsunpcSq$, while the combined WNM has a combined surface density of $4.65\, \MsunpcSq$. The errors are assumed to be dependent and so are added linearly, producing a combined uncertainty of 15\%. However, \cite{McKee} earlier stated that the fiducial Heiles model \citep{Heiles1981} for optically thin HI (which includes the WNM) is accurate to within 20\%. Given this potential ambiguity we decided to take 20\% as the uncertainty for all HI components. 

\cite{2017arXiv170700743M} recently presented a new determination of the HI surface density of $\Sigma^\infty_{\rm HI} = 4.5 \pm 0.7\, \MsunpcSq$, or $\Sigma^\infty_{\rm HI} = 6.3 \pm 1.0\, \MsunpcSq$ when including the 1.4 helium correction factor.
If this figure is correct the total baryon surface density would be lowered by $\sim 10\%$. Using the two dimensional posterior from Fig. \ref{rhoDM_SigBa_tilt_fig} as a guide we estimate that this would raise the $\rhodm$ determination by $\sim 17\%$. 

\subsubsection{Ionized gas HII surface density}
For the HII ionized gas surface density uncertainty we return to the several of the works cited by \cite{McKee}. The ionized gas consists mainly of warm photo-ionized gas at temperature $\sim 10^4$~K and hot collisionally ionized gas at $\sim 10^5 - 10^6$~K, where the latter only gives a small contribution in our region of interest and is hence not treated here \citep{Gaensler, McKee}. The photo-ionized gas in the solar vicinity is dominated by a diffuse thick disc and the Gum Nebula; the latter is estimated to contribute about 0.11~M$_\odot$~pc$^{-2}$ \citep{McKee} to the total surface density. Here we will treat the Gum Nebula contribution as localized to the disc plane.

The scale height of the diffuse thick disc of ionized gas has in more recent work been found to have a larger scale height than previously believed, for example in \cite{Gaensler} they find a scale height of $1830^{+120}_{-250}$~pc. The recent review by \cite{Schnitzeler} finds that the best fit is achieved by an exponential thick disc with a scale height of 1590~kpc and a column number density perpendicular to the plane of $24.4\pm 4.2$~cm$^{-3}$~pc, which corresponds to a surface number density of $1.5 \pm 0.26 \times 10^{20}$~cm$^{-2}$. Hence we arrive at a surface density of $1.7 M_{\odot}$pc$^{-2}$ for the $HII$ diffuse thick disc, and $1.8 M_{\odot}$pc$^{-2}$ when including the Gum Nebula; the former is used in the $HII$ profile in Figs.~\ref{barydens_fig} and \ref{barysurfdens_fig}, and the latter is used for the total baryonic surface density (at $z\to\infty$). The surface number density uncertainty corresponds to an error of 17\%, higher than the 6\% quoted in table 2 of \cite{McKee} et al. In the face of this discrepancy we err on the conservative side and take 17\% as our uncertainty on the HII surface density.
 
\subsubsection{Combination of surface density uncertainties}
In deriving the combined uncertainties on the baryon surface densities we make the assumption that the gas component uncertainties are dependent with each other, the stellar component uncertainties are likewise dependent on each other, but the total gas uncertainty is independent from the total stellar uncertainty. This means that within each category (gas and stars) the errors are added linearly, but the addition of gas uncertainty to stellar uncertainty is done in quadrature. We make these assumptions because it is unclear that the gas component uncertainties are independent, and similarly the stellar component uncertainties, thus leading to the conservative assumption that they are indeed correlated. 
 
\subsection{Vertical profile modelling}\label{sec:baryon_vertical_profile_modelling}
As mentioned earlier the HII gas modelling is complicated by the presence of the Gum Nebula. The profile we take is an exponential with a scale height of 1.59 kpc, and the surface density used to normalise this profile is $\Sigma_{\rm HII}^\infty = 1.7~\MsunpcSq$ (\eg without the contribution of the Gum Nebula, see above). 

The two stellar surface density profiles we model are the thin disc component of the dwarfs, and a profile that sums the thick disc components of the MS stars and the dwarfs. Each of these populations will have thick and thin disc components; here the vertical profile of the MS thin disc component is not modelled, and the thick disc vertical profiles of the MS stars and dwarfs are modelled jointly. The procedure is to first construct a vertical profile featuring a thin and thick disc, but no MS-dwarf partition. Then we calculate the fraction of dwarfs in the thick disc, and the same for the MS stars. The sum of the surface densities of MS and dwarf thick disc components is then used as the surface density for the total thick disc vertical profile. 

The total density profile for a population featuring a thick disc can initially be described by 
\begin{equation}
\rho(z) = \rho_0 \left[(1-\beta)\mathrm{sech}^2\left(\frac{z}{h_1}\right) + \beta\exp\left(-\frac{z}{h_2}\right)\right],\label{rho_thin_thick}
\end{equation}
where $\rho_0$ is the density of stars at $z=0$, $h_2 > h_1$, and the $\beta$ parameter is the proportion of $\rho_0$ attributed to the thick disc (\ie $\beta \in [0,1]$). Integrating Eq. \ref{rho_thin_thick} to infinity gives the surface density, \ie:
\begin{equation}
\Sigma = 2\int_0^\infty \rho\,\mathrm{d}z = 2\rho_0[(1-\beta)h_1+\beta h_2] = 2\rho_0h_{\rm eff, tot},
\end{equation}
where
\begin{equation}
h_{\rm eff, tot} \equiv \frac{\Sigma}{2\rho_0} = (1-\beta)h_1+\beta h_2 \label{eq:h_expr},
\end{equation}
is the effective scale height for the total distribution. 

From equation (\ref{eq:h_expr}) one sees that for a given value of $h_{\rm eff, tot}$ it is possible to either have a thick disc with a large scale height, $h_2$, and small contribution to the local density at the disc plane $\beta$; or a thick disc with a small $h_2$ and a large $\beta$.  For M dwarfs \cite{Z01} have quite large $\beta$ and small $h_2$, while \cite{F06} model all thick discs with a small $\beta$ and large $h_2$. There is a priori no reason why these thick disc components should be modelled differently, and both models seem reasonable for all stars. To encompass this freedom we expand the thick disc model to a sum of two thick discs: one with a slightly smaller scale height $h_2$ and one with a larger scale height $h_3$. The density profile of a stellar component with a thick disc is hence modelled as
\begin{equation}
\begin{split}
\rho(z) = \rho_0\left[(1-\beta)\mathrm{sech}^2\left(\frac{z}{h_1}\right) + \beta (1-x)\exp\left(-\frac{z}{h_2}\right) \right. \\
+\left. \beta x\exp\left(-\frac{z}{h_3}\right)\right], \label{eq:thin_doublethick}
\end{split}
\end{equation}
where the first term is the thin disc component, $\rho_0$ is again the local density at $z=0$, and $x$ is the relative contribution of the two thick disc components to $\rho_0$, i.e. a positive number that cannot be greater than one. We can assign the thick disc an effective scale height: $h_{\rm eff, thick}=(1-x)h_2 + xh_3$, so that we can still make use of relation (\ref{eq:h_expr}): 
\begin{equation}
h_{\rm eff, tot}=(1-\beta)h_1+\beta h_{\rm eff, thick},\label{beta_eq}
\end{equation}
which then gives
\begin{equation}
\beta = \frac{h_{\rm eff, tot}-h_1}{h_{\rm{eff, thick}}-h_1}.
\label{eq:beta_as_hfractions}
\end{equation}
For the modelling of thick disc we require a hierarchy of scale heights: $h_1<h_{\rm eff, tot}<h_2<h_3$; from the expression for $h_{\rm eff,thick}$ one then gets that $h_2\leq h_{\rm eff, thick}\leq h_3$, which in turn automatically implies that $0<\beta <1$. This model of the thick disc is much less constrictive than that of \cite{McKee}, and so better reflects the uncertainty present in the literature. 

The scale height for the dwarf thin disc component (we do not model the MS thin disc component) is set to $h_1=332$~pc~$\pm 10\%$, drawing from \cite{Z01}. The parameters we set for these two thick disc components are $h_2=609$~pc~$\pm 20\%$ taken from \cite{Z01} (assuming the Color-Magnitude Relation (CMR) 1 option from that paper), $h_3 = 1000$ pc $\pm 20\%$ from table 2 of \cite{F06} (combining $\rho(z=0) = 0.0035\, \Msunpc$ and $\Sigma = 7.0\, \MsunpcSq$ to yield $h = \Sigma/2\rho_0 = 1000$ pc), and $h_{\rm eff, tot} = 400$~pc~$\pm 5\%$ \citep{McKee} (relevant for only the dwarf population). We have increased the uncertainties on the numbers from \cite{Z01} in an attempt to encompass uncertainties arising from different CMRs. The size of the uncertainties are set so they do not overlap and violate the scale height hierarchy described above. The value of $x$ is taken from a flat prior between 0 and 1.

The total surface density of the thick disc described here is a combination of the thick disc components of the dwarfs and MS stars. We thus need to first calculate the partitioning of each component into thick and thin pieces, using the $\beta$ parameter. The $\beta_{\rm dwarf}$ parameter is calculated as per equation (\ref{eq:beta_as_hfractions}), yielding
\begin{equation}
\Sigma_{\rm dwarf,thick}^\infty = \Sigma_{\rm dwarf}^\infty\, \beta_{\rm dwarf}\, \frac{h_{\rm eff, thick}}{h_{\rm eff, tot}}. 
\label{eq:Sigma_dwarf_thick}
\end{equation}
The choice of $h$ parameters above yields $0.018 \leq \beta_{\rm dwarf} \leq 0.64$, and the total surface density of dwarfs is taken to be $\Sigma_{\rm dwarf} = 23.7 \MsunpcSq \pm 20$\%, where $17.3 \MsunpcSq$ comes from M dwarfs and the rest is from the stellar remnants \citep{McKee}. 

The portioning of the MS star surface density uses $\beta_{{\rm MS,\,thick}}$ taken from a flat prior between 0 and 0.3. Thus
\begin{equation}
\Sigma_{{\rm MS,\,thick}}^\infty = 2\, h_{\rm eff, tot} \,\rho_{0,{\rm MS,\,thick}},
\label{eq:Sigma_MS_thick}
\end{equation}
where 
\begin{equation}
\rho_{0,\,{\rm MS,\,thick}} = \rho_{0,{\rm MS}} \, \beta_{{\rm MS}}.
\label{eq:rho0_MS_thick}
\end{equation}
The density at $z=0$ is $\rho_{0,{\rm MS}} = 0.0107 \, \Msunpc$ \citep[MS stars with $4<M_V<8$ plus giants,][]{McKee}. This number is taken without uncertainty as it is actually $\rho_{0,\,{\rm MS,\,thick}}$ which is used for the baryon modelling, and this has sufficient freedom coming from the $\beta_{{\rm MS}}$ parameter.

\subsection{Contribution to $\sigma_z^2$ by a potentially undetected baryonic thick disc}
In modelling the baryonic mass distribution there is always a risk that we miss some low density baryonic component that has a very large scale height and hence might still be important at high $z$ where we have less data. Such a baryonic thick disc component is typically expected to have have a shape of the form $\rho(z)=\rho(0)\mathrm{e}^{-z/\ell}$ where $\ell$ is its scale height. As before our tracer stars are well approximated by $\nu(z)\propto\mathrm{e}^{-z/h}$; the contribution to the tracer velocity dispersion from this unrecognized baryonic thick disc component then becomes
\begin{equation}
\frac{\sigma_z^2(z)}{2\pi Gh} = \frac{h}{h+\ell}\Sigma(\infty) + \frac{\ell}{h+\ell}\Sigma(z). \label{bary_thick_not_imp}
\end{equation}
The density contribution outside the point of investigation looks the same to $\sigma_z^2$ as a surface density component confined to the disc plane. The contribution from $\Sigma(\infty)$ in equation (\ref{bary_thick_not_imp}) is small compared to the contribution from $\Sigma(z)$; both because $h/(h+\ell)\ll \ell/(h+\ell)$ since a thick disc per definition has a large scale height $\ell \gg h$ , and because the total surface density of a neglected baryonic component is expected to be smaller than \eg $\Sigma_{b0}$ of equation~(\ref{sigz2_not_tilt2_eq}). Hence the importance of an unrecognized baryonic thick disc is mainly evaluated by its contribution to the total surface density at the $z$ region where we have stellar tracer data, which is also a region where we expect to have the baryonic contributions decently under control.

\section{Results}\label{sec:results}

\subsection{Simplified analytic analysis}\label{simplified_analytic_analysis_section}
\begin{figure}
\includegraphics[scale=0.475, trim={0.7cm 0.2cm 2.0cm 1.25cm},clip]{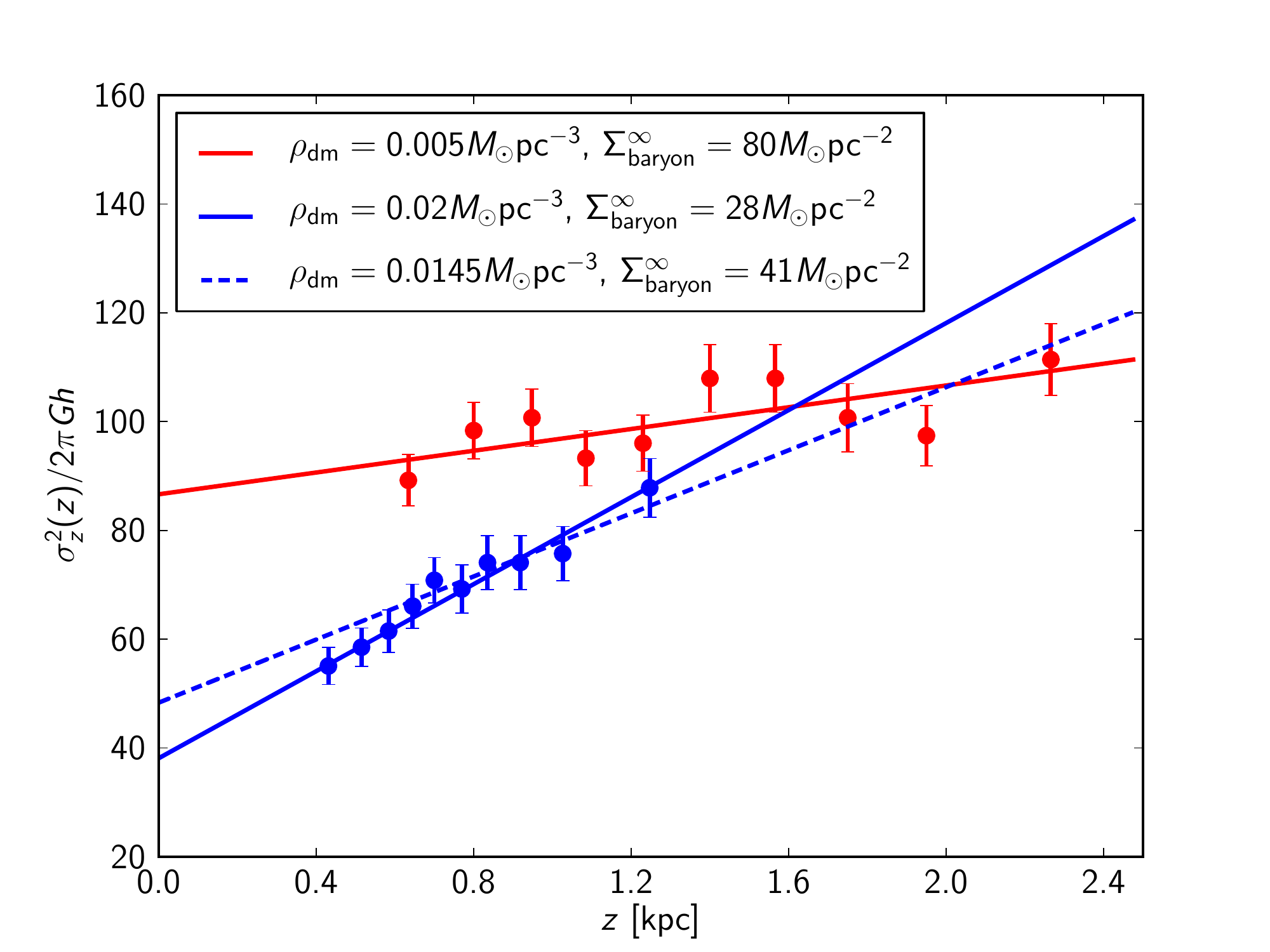}  
\caption{The tilt-free simplified modelling of equation~(\ref{analytic2}) (solid lines), fitted to and plotted with $\sigma_z^2$ data divided by $2\pi Gh$ for the $\alpha$-young (blue, lower) and $\alpha$-old (red, upper) populations. Here we assume that the tracer densities for the two populations each consist of a single exponential with scale height $h_{\rm young}=253$~pc and $h_{\rm old}=665$~pc for the $\alpha$-young and $\alpha$-old populations, respectively, which are the fits of B15. As seen in the plot the two populations do not prefer the same values on $\rho_{\rm{dm}}$ and $\Sigma^{\infty}_{\rm{baryon}}$, and neither of the solid line fits agree with the total baryonic surface density of Section \ref{sec:baryon_modelling}: $\Sigma_{\rm{baryon}}^\infty = 46.85$~$M_\odot$pc$^{-2}\pm 13\%$. However, the approximation that all the baryonic mass is inside the innermost bin is not entirely true, especially for the $\alpha$-young data. If we take this into account and allow the fit to overshoot the low $z$ data points we can, for the $\alpha$-young data, instead make the blue dashed fit which has a slightly lower dark matter density than the blue solid line and a baryon surface density in agreement with the result of Section \ref{sec:baryon_modelling}. We cannot, however, play the same trick with the $\alpha$-old data and this gives a slight tension, further discussed in the text. Also recall that the $\alpha$-old data is affected more by the tilt term than the $\alpha$-young data.}
\label{simple_sigz2_fit_fig}
\end{figure}

To gain some intuition on the analysis we investigate the simplified case where all baryons are inside $z_-$, \ie in equation~(\ref{sigz2_not_tilt2_eq}) set $\Sigma_{b1}(z)\to 0$. As we see in Fig.~\ref{barysurfdens_fig} this approximation is not very far from the truth. For this simplified case where we also disregard the tilt term, we then get:
\begin{equation}
\frac{\sigma_z^2(z)}{2\pi Gh} = \Sigma_{\mathrm{baryon}}^\infty + 2\rho_{\mathrm{dm}}h + 2\rho_{\mathrm{dm}}z {\rm \ \ \ \ \ for \ } z>z_- \label{analytic2},
\end{equation}
where $\Sigma_{\mathrm{baryon}}^\infty$ is the total baryonic surface density at infinity. In equation (\ref{analytic2}) the slope of $\sigma_z^2(z)/(2\pi Gh)$ is directly determined by $\rho_{\mathrm{dm}}$. Hence when neglecting the tilt term the slopes of $\sigma_z^2(z)/(2\pi Gh)$ should be the same for the two tracer populations since they both probe the same matter density distribution; this observation is true independent of baryonic mass distribution. After measuring the slope, \ie measuring $\rho_{\mathrm{dm}}$, $\Sigma_{\mathrm{baryon}}^\infty$ can then be determined by looking at $\sigma_z^2(z\to 0)/(2\pi Gh)$. 

We can then fit this simplified model by eye to the $\alpha$-young and $\alpha$-old $\sigma_z^2(z)$ data from B15 used throughout this paper. In Fig. \ref{simple_sigz2_fit_fig} we plot these fits (solid blue and red for $\alpha$-young and $\alpha$-old respectively), scaling $\sigma_z^2(z)$ by $(2\pi G h)^{-1}$ to match with equation (\ref{analytic2}), assuming scale heights of  $h_{\rm young}=253$~pc and $h_{\rm old}=665$~pc. These fits show that in the simplified analysis the $\alpha$-young and $\alpha$-old populations prefer different values for baryonic surface density and dark matter density: the $\alpha$-young population prefers $\rhodm = 0.02\, \Msunpc = 0.76\, \gevcm$ and $\Sigma_{\rm baryon}^\infty = 28\, \MsunpcSq$, while $\alpha$-old prefers a lower DM density of $\rhodm = 0.005\, \Msunpc = 0.19\, \gevcm$ and a higher baryon surface density of $\Sigma_{\rm baryon}^\infty = 80\, \MsunpcSq$. This is similar to B15 where they also find the two populations to prefer different dark matter densities. Also note that the preferred baryon surface densities of both populations, but particularly the $\alpha$-old population, are outside the range we derive from baryon census measurements of $\Sigma_{\rm{baryon}}^\infty = 46.95$~$M_\odot$pc$^{-2}\pm 13\%$ (see Section \ref{sec:baryon_modelling}). 

The approximation that all baryon density is confined to the mid-plane at $z=0$, \ie below the range of the data, is especially stretched for the lower data points of the $\alpha$-young population. We can investigate the impact of this assumption by neglecting the lower three data points from the fit. This produces the dashed blue line in Fig. \ref{simple_sigz2_fit_fig}, and results in a preferred baryon surface density of $\Sigma_{\rm baryon}^\infty = 41\, \MsunpcSq$, in agrement with the figure we derive in Section \ref{sec:baryon_modelling}, and a DM density of $\rhodm = 0.0145\, \Msunpc = 0.55\, \gevcm$. Given the higher starting point of the $\alpha$-old data, confining the baryons to the mid-plane is a more valid assumption, and so this assumption is unlikely to be the cause of the discrepancy between the $\alpha$-old preferred $\Sigma_{\rm baryon}^\infty$ and the baryon census derived result. 

We must also investigate the impact of the tilt term on this example. The tilt term contribution is expected to be more important for the $\alpha$-old population than for the $\alpha$-young population. The tilt term is expected to be zero at $z=0$, and increase in magnitude with increasing $z$. This is a similar behaviour to the DM surface density $\Sigma_{\rm dm} = \rhodm z$, and so the tilt term can easily change the derived DM density, either increasing or decreasing it depending on the radial behaviour of the tilt term. Thus the tilt term could, if large enough, make the two populations in Fig.~\ref{simple_sigz2_fit_fig} agree on the dark matter density. However, as the tilt term is zero at $z=0$ and generally smaller at small $z$, it would be more difficult for it to account for the discrepancy in preferred $\Sigma_{\mathrm{baryon}}^\infty$ between $\alpha$-young and $\alpha$-old, given that the baryons are confined to $z=0$ in our example, or at least heavily concentrated at low-$z$ in reality. The discrepancy between the $\alpha$-young and $\alpha$-old populations is discussed further in Section \ref{sec:young_old_discrepancy}.

\subsection{Full \MultiNest Analysis}\label{sec:results_full_analysis}

\begin{table*}
\caption{Summary of results for the credible region (CR) of the marginalised posterior for $\rhodm$, for separate $\alpha$-young and $\alpha$-old analyses (with tilt and without), and for a combined $\alpha$-young and $\alpha$-old analysis.The most reliable result if from the $\alpha$-young with tilt analysis, shown in bold face.}
\label{tab:results}
\begin{tabular}{l l | c c | c c | c}
\hline
&& \multicolumn{2}{c}{$\alpha$-young} & \multicolumn{2}{c}{$\alpha$-old} & Combined analysis \\
&& Tilt & No Tilt & Tilt & No Tilt & Tilt \\
\hline
\hline
95\% CR upper & $\gevcm$ & $\mathbf{0.59}$ & $0.57$ & $0.85$ & $0.51$ & $0.48$\\
& $\Msunpc$ & $\mathbf{0.016}$ & $0.015$ & $0.022$ & $0.013$ & $0.013$\\
&&&&&&\\
68\% CR upper & $\gevcm$ & $\mathbf{0.53}$ & $0.53$ & $0.79$ & $0.48$ & $0.43$\\
& $\Msunpc$ & $\mathbf{0.013}$ & $0.014$ & $0.021$ & $0.013$ & $0.012$\\
&&&&&&\\
Median & $\gevcm$ & $\mathbf{0.46}$ & $0.48$ & $0.73$ & $0.46$ & $0.40$ \\
& $\Msunpc$ & $\mathbf{0.012}$ & $0.013$ & $0.019$ & $0.012$ & $0.011$ \\
&&&&&&\\
68\% CR lower & $\gevcm$ & $\mathbf{0.37}$ & $0.42$ & $0.68$ & $0.44$ & $0.37$\\
& $\Msunpc$ & $\mathbf{0.0098}$ & $0.011$ & $0.017$ & $0.012$ & $0.0097$\\
&&&&&&\\
95\% CR lower & $\gevcm$ & $\mathbf{0.30}$ & $0.35$ & $0.60$ & $0.42$ & $0.34$ \\
& $\Msunpc$ & $\mathbf{0.0078}$ & $0.0092$ & $0.016$ & $0.011$ & $0.0091$\\
\hline
\end{tabular}
\end{table*}

\begin{figure*}
\includegraphics[width=1.0\textwidth, trim={0 9cm 0 0},clip]{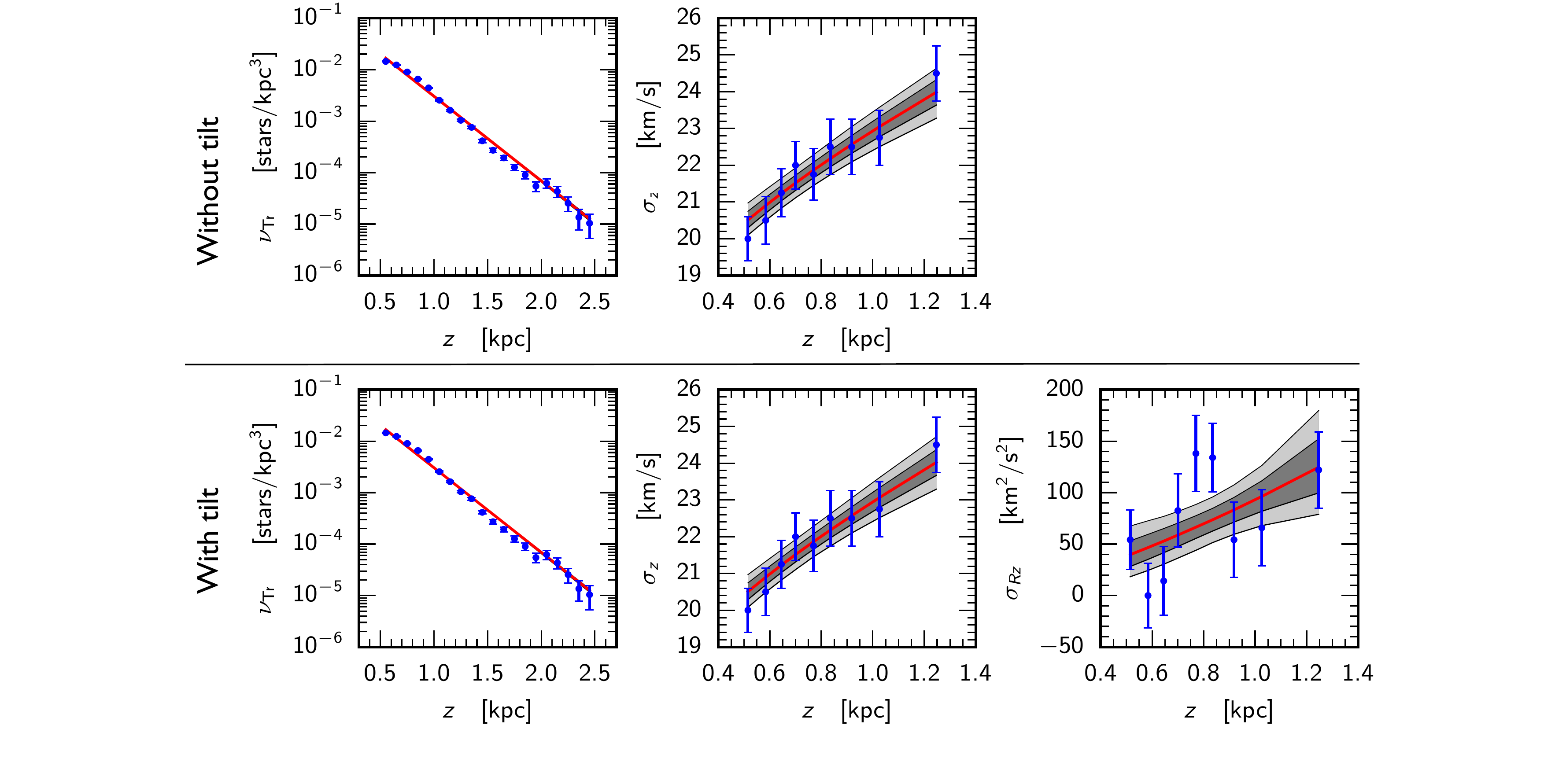}\\   
\includegraphics[width=1.0\textwidth, trim={0 0.6cm 0 0},clip]{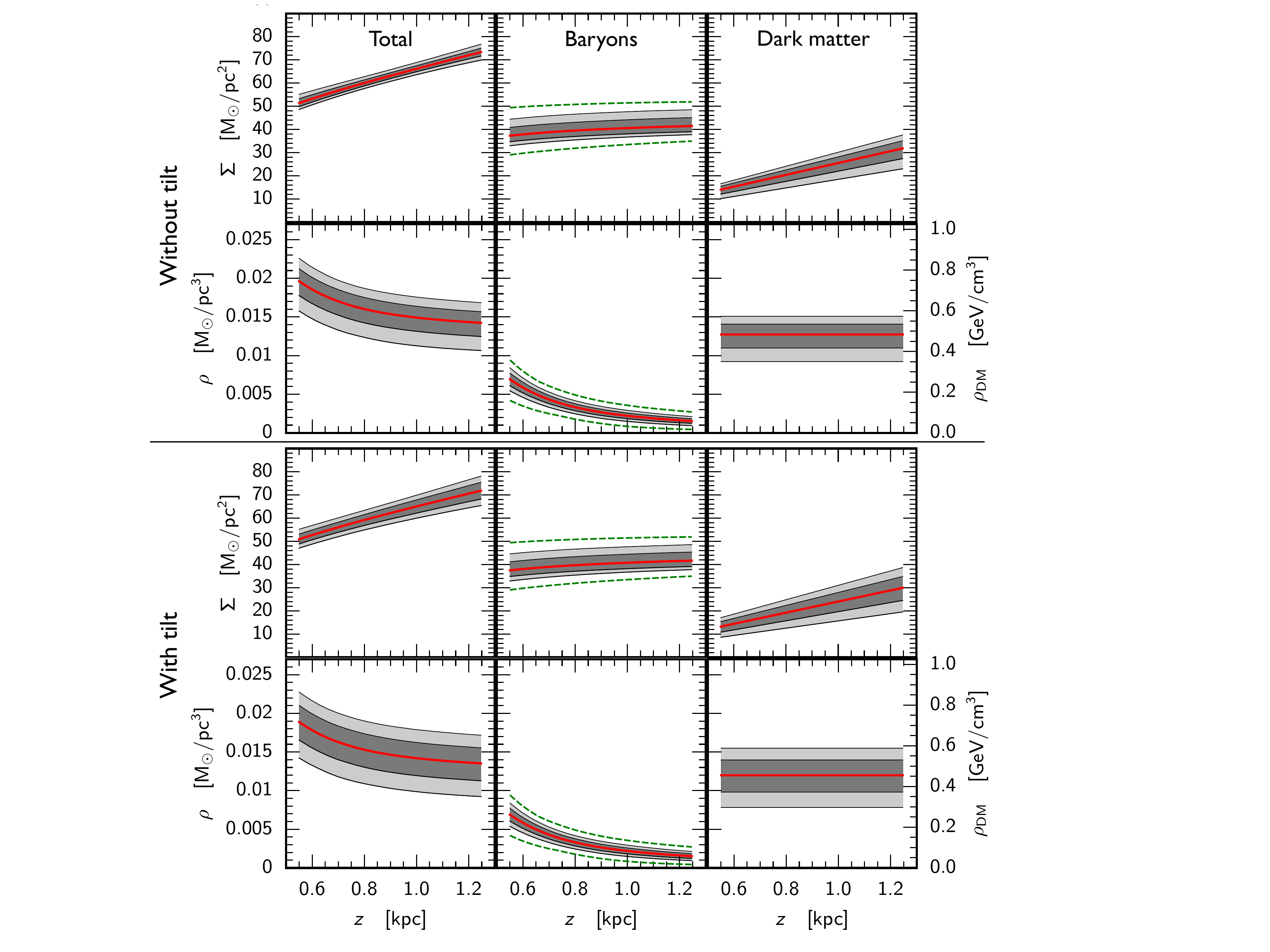} 
\caption{Result of analysing the $\alpha$-young data from, with and without taking the tilt term into account. The blue points are the data from B15 to which the model is fitted. Dark and light gray shaded regions show the 68\% and 95\% credible regions, respectively; the red lines show the median values. Green dotted lines show the prior range of the baryonic model. The resulting dark matter density is 0.46 $\gevcm$ (68\% CR: [0.37, 0.53] $\gevcm$, 95\% CR: [0.30, 0.59] $\gevcm$) or 0.012 $\Msunpc$ (68\% CR: [0.0098, 0.013] $\Msunpc$, 95\% CR: [0.0078, 0.016] $\Msunpc$) when taking the tilt term into account, and 0.48 $\gevcm$ (68\% CR: [0.42, 0.53] $\gevcm$, 95\% CR: [0.35, 0.57] $\gevcm$) or 0.013 $\Msunpc$ (68\% CR: [0.011, 0.014] $\Msunpc$, 95\% CR: [0.0092, 0.015] $\Msunpc$) when not taking tilt into account.}
\label{young_pop_fig}
\end{figure*}

\begin{figure*}
\includegraphics[width=1.0\textwidth, trim={0 9cm 0 0},clip]{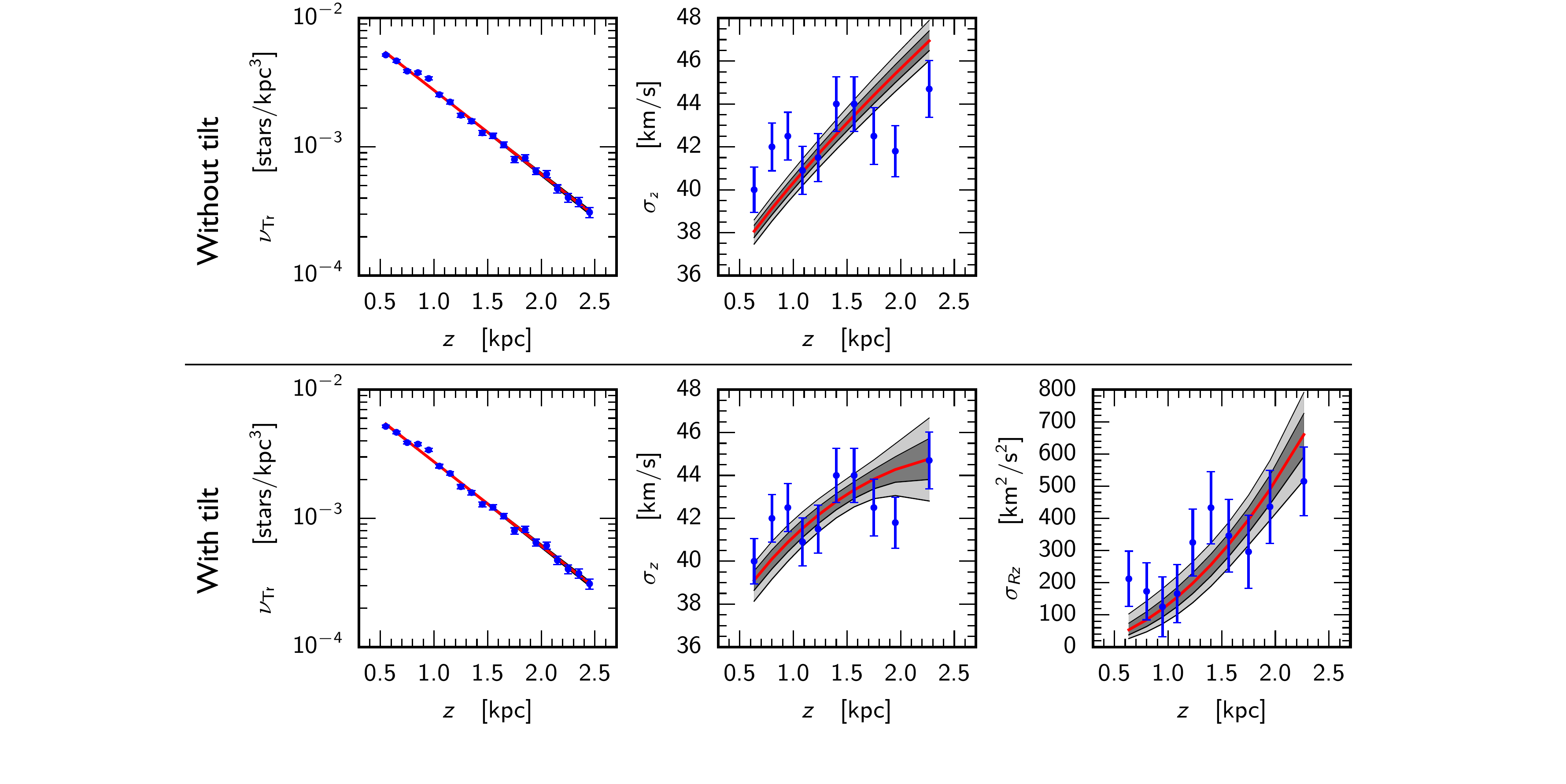}\\   
\includegraphics[width=1.0\textwidth, trim={0 0.7cm 0 0},clip]{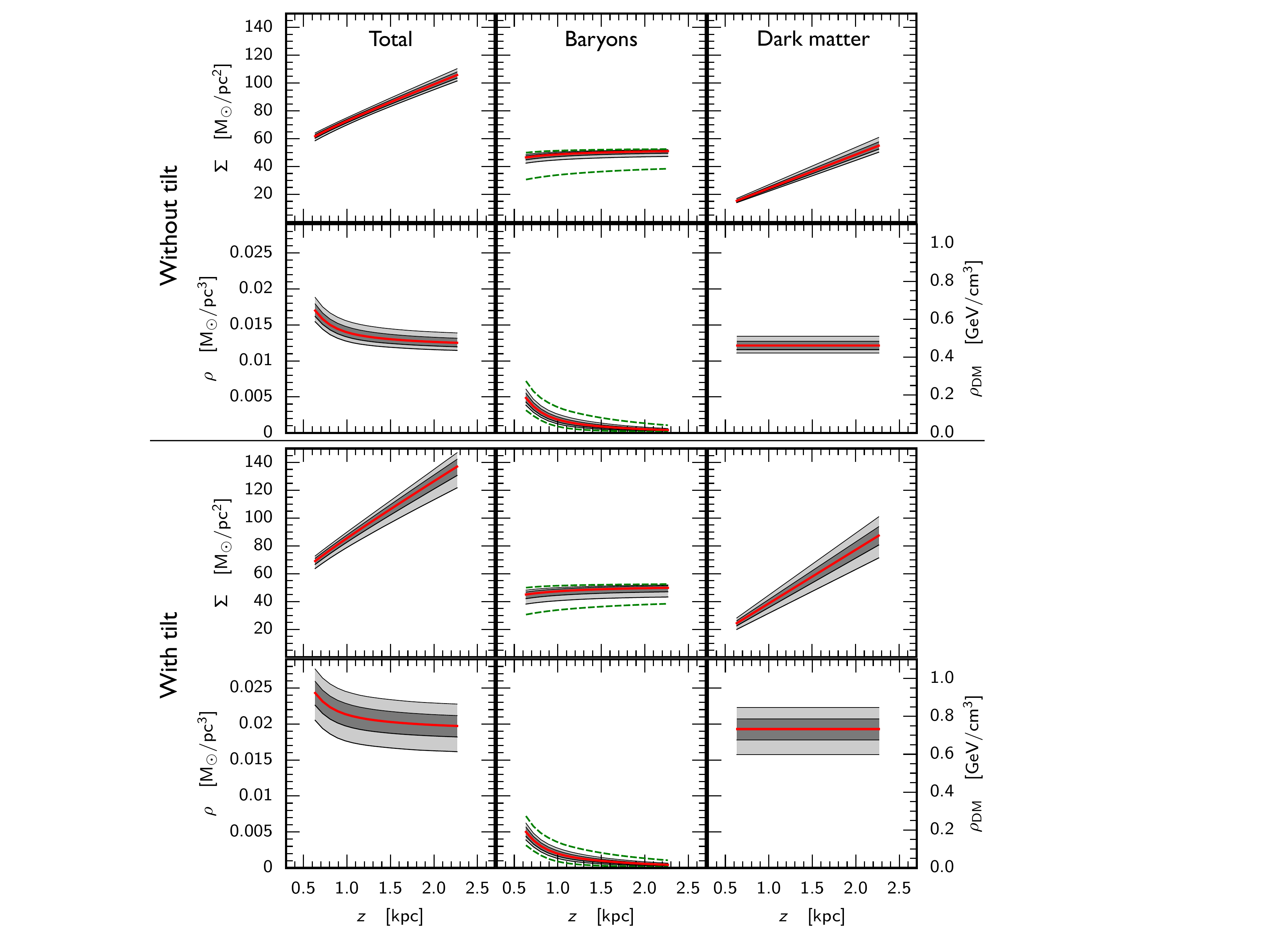} 
\caption{Results of the analysis of the $\alpha$-old data from B15, with and without taking the tilt term into account in the analysis. The lines, points, and shaded regions are as in Fig.~\protect\ref{young_pop_fig}. The resulting dark matter density is 0.73 $\gevcm$ (68\% CR: [0.68, 0.79] $\gevcm$, 95\% CR: [0.60, 0.85] $\gevcm$) or 0.019 $\Msunpc$ (68\% CR: [0.017, 0.021] $\Msunpc$, 95\% CR: [0.016, 0.022] $\Msunpc$) when taking the tilt term into account, and 0.46 $\gevcm$(68\% CR: [0.44, 0.48] $\gevcm$, 95\% CR: [0.42, 0.51] $\gevcm$) or 0.012 $\Msunpc$ (68\% CR: [0.012, 0.013] $\Msunpc$, 95\% CR: [0.011, 0.013] $\Msunpc$) when not taking tilt into account. Note how large impact the inclusion of the tilt term has on the recovered dark matter density, presumably a result of the poor fit to the $\sigma_{z}$ data for the analysis neglecting tilt. Also note that the fit to the $\sigma_{Rz}$ data is not excellent for the analysis including the tilt term.}
\label{old_pop_fig}
\end{figure*}

\begin{figure*}
\includegraphics[width=0.98\textwidth, trim={0 9cm 0 0},clip]{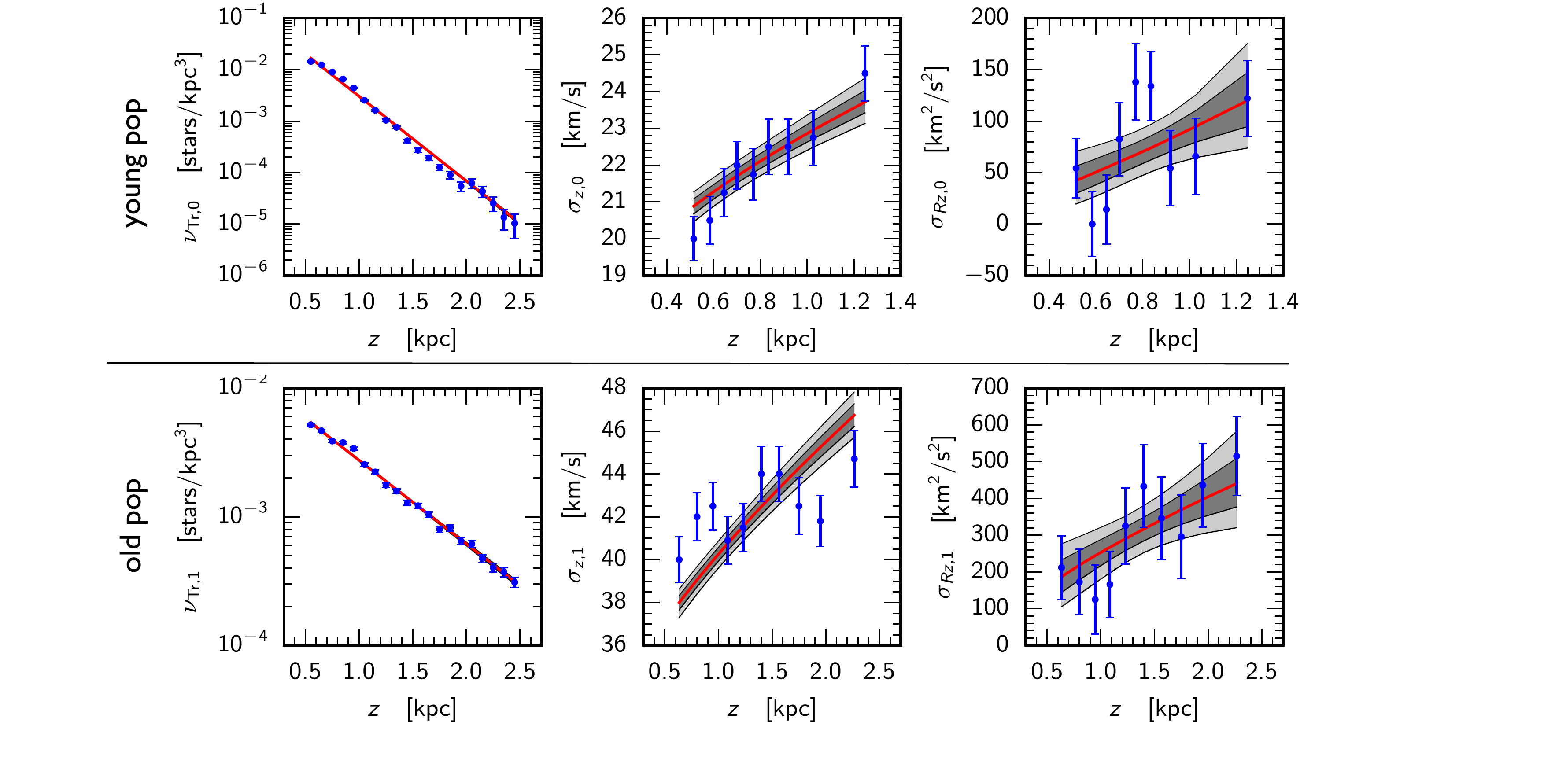}\\   
\includegraphics[width=0.98\textwidth, trim={0 12.2cm 0 0},clip]{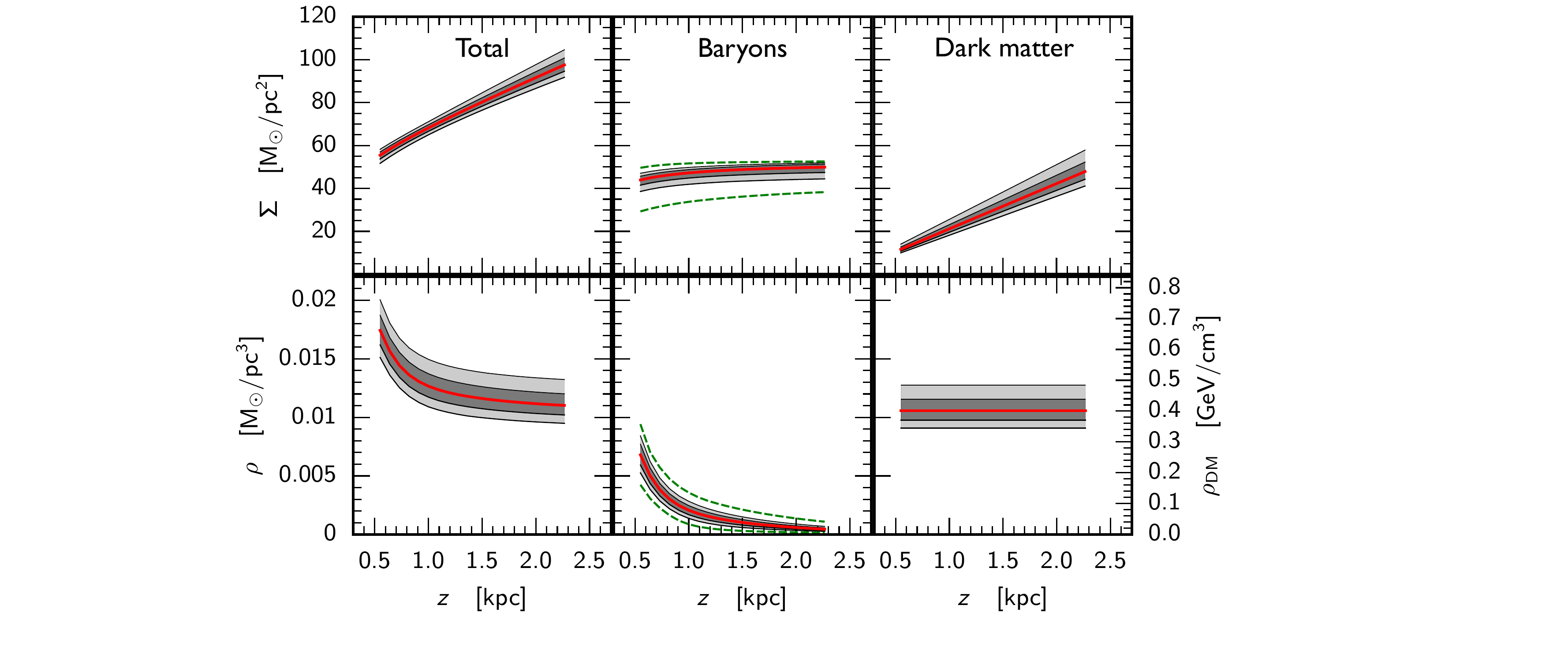} 
\caption{Result of a combined analysis of data for the $\alpha$-young and $\alpha$-old populations from B15, tilt modelling included. The convention is the same as in Fig.~\protect\ref{young_pop_fig} and \protect\ref{old_pop_fig}. The resulting dark matter density is 0.40 $\gevcm$ (68\% CR: 0.37, 0.43 $\gevcm$, 95\% CR: 0.34, $\gevcm$) or 0.011 $\Msunpc$ (68\% CR: 0.0097, 0.012 $\Msunpc$, 95\% CR: 0.0091, 0.013 $\Msunpc$). Note how much more similar the resulting dark matter density is to the with tilt fit to the $\alpha$-young data (Fig.~\ref{young_pop_fig}), than to that of the $\alpha$-old data (Fig.~\ref{old_pop_fig}).}
\label{2pop_fig}
\end{figure*}

Here we now present the results of five different analyses using the full \MultiNest analysis: $\alpha$-young only, with and without tilt; $\alpha$-old only, with and without tilt, and a combined $\alpha$-young and $\alpha$-old analyses with tilt. The results of these analyses are summarised in Table~\ref{tab:results}, which gives the values for the median, 95\% and 68\% credible region bounds on the marginalised posterior for $\rhodm$. The plots shown in Figs.~\ref{young_pop_fig}, \ref{old_pop_fig}, and \ref{2pop_fig} display the $z$-profiles of a number of quantities. Data points and uncertainties are marked as blue points and error bars, while the light grey bands, dark grey bands, and red lines indicate the 95\% limits, 68\% limits, and median values of the posterior distribution. For the baryon distribution plots the dotted green lines show the limits of the prior range of the baryonic model.   

When fitting only the $\alpha$-young data, shown in Fig. \ref{young_pop_fig}, the model manages to fit the data well despite not taking the tilt term into account, something seen earlier in the simplified analysis of Section \ref{simplified_analytic_analysis_section}. This result is as expected since the $\alpha$-young data set consists of a colder stellar population which is more confined to the disc plane, where we do not expect the tilt term to have a large impact as discussed at the end of Section~\ref{K_prior} and \eg~ B15. As per expectation, the inclusion of the tilt term in the analysis of the $\alpha$-young data only has a minor impact on the recovered dark matter density: without tilt the median $\rhodm = 0.013\,\Msunpc = 0.48\,\gevcm$, while with tilt it is $\rhodm = 0.012\,\Msunpc = 0.46\,\gevcm$. Inclusion of the tilt term mainly results in a widening of the error bands, as can be seen by comparing the with tilt and without tilt results of Fig.~\ref{young_pop_fig}.

The tilt term is expected to be more important for the analysis of the $\alpha$-old data. This is a older, hotter stellar population which reaches further above the disc plane. In Fig.~\ref{old_pop_fig} we see that for the analysis of the $\alpha$-old data without tilt, the fit to the $\sigma_z$ data is not very good: the recovered $\sigma_z$ model increases more quickly with $z$ than the data points. We also see the baryon surface density is concentrated against the upper end of the prior imposed by baryon observations, reflecting the preference the $\alpha$-old data displayed for high baryon surface density in our earlier simplified analysis (Section \ref{simplified_analytic_analysis_section}). This tension in the data when neglecting the tilt term results in an overly constrained recovered dark matter density, as seen in Fig. \ref{old_pop_fig}. For this analysis the median of the marginalised DM density posterior is $\rhodm = 0.012 \Msunpc = 0.46\,\gevcm$.  

Including the tilt term in the analysis of the $\alpha$-old data results in a somewhat better fit to the $\sigma_z$ data; this is however accomplished by using a high value on the tilt parameter $n$, giving a rather steep fit to the $\sigma_{Rz}$ data, as seen in Fig. \ref{old_pop_fig}. The recovered dark matter density for the analysis of the $\alpha$-old data has a more reasonably sized credible region, but one that has moved significantly upwards: the median DM density for this analysis is $\rhodm = 0.019 \Msunpc = 0.73\,\gevcm$, and the 95\% CRs of the with tilt and without tilt analyses do not overlap. Furthermore the 95\% CRs of the $\alpha$-young and $\alpha$-old analyses with tilt do not overlap. 

We can also perform a combined fit to both the $\alpha$-young and $\alpha$-old population, using common DM and baryon distributions, but with separate tracer density and tilt profiles for each population. This joint analysis, including the tilt term, is shown in Fig. \ref{2pop_fig}. This joint analysis results in a dark matter density CR which is compatible with the $\alpha$-young results (with and without tilt) at the 68\% level, and compatible at the 95\% level with the $\alpha$-old without tilt result. The 95\% CRs of the joint and $\alpha$-old with-tilt analysis do not overlap. 

The with-tilt analyses of the $\alpha$-young and $\alpha$-old populations are incompatible at the $95\%$ level or above, and yet the joint analysis, which includes tilt, favours the $\alpha$-young population. The tendency of the $\alpha$-old data to favour higher baryon surface densities persists in the joint analysis, most likely causing the reduced $\rhodm$ result of the joint analysis compared to the $\alpha$-young with tilt result. 

Due to the poorness of the fit to the $\sigma_z$ data for the 2 population joint analysis, shown in Fig.~\ref{2pop_fig}, and concerns with the $\alpha$-old data which will be discussed in Section \ref{sec:young_old_discrepancy}, for our final result we will instead use the results of the $\alpha$-young with-tilt analysis, as shown in Fig.~\ref{young_pop_fig}. We emphasize that this choice does not have a outsized impact on the reported dark matter density as the $\rhodm$ CR of the joint analysis is contained with that of the $\alpha$-young with-tilt analysis.

\subsection{Degeneracy between the dark matter density and the baryonic surface density}\label{sec:dm_baryon_degeneracy}

\begin{figure}
\includegraphics[scale=0.73, trim={0.2cm 0.27cm 0 0.2cm},clip]{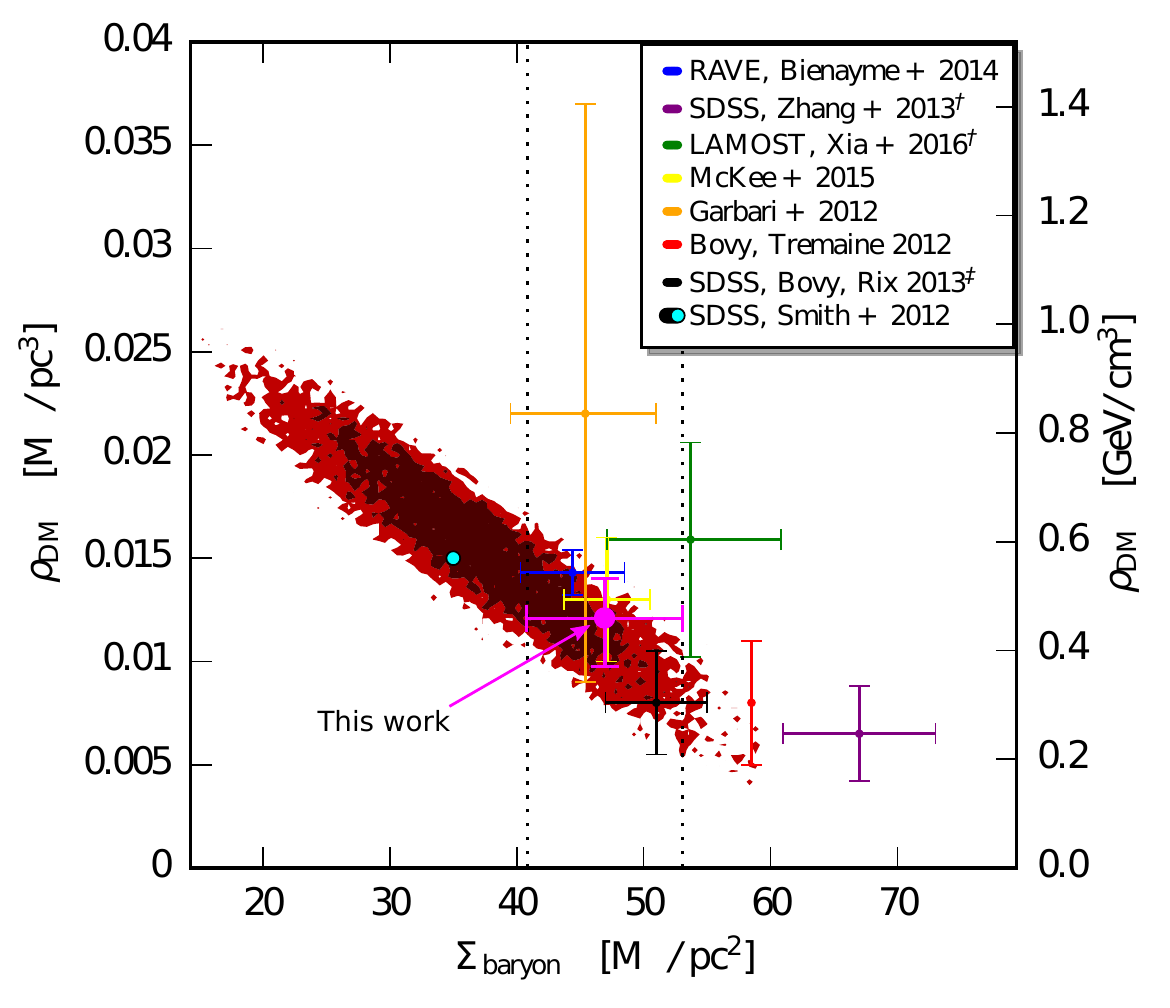} 
\caption{Two-dimensional marginalised posterior for the local dark matter density and the baryonic surface density, illustrating the degeneracy between the two. This posterior is the product of a \MultiNest run with significantly expanded baryon surface density priors, so as to better show the degeneracy and its relation with previous local DM results, plotted as points and errorbars. The normal baryon surface density priors used for all other runs throughout this paper are shown by the vertical dotted lines (c.f. Section \protect\ref{sec:baryon_modelling}). The input model for the two-dimensional marginalised posterior includes tilt, and the input data is that of the $\alpha$-young population. The definitive $\rhodm$ result for this work, derived from the $\alpha$-young with tilt analysis, is shown in magenta. Neglecting the tilt term would result in the red ellipses appearing with the same orientation and major axes, but with contracted minor axes. Plotted are also recent results from other authors; for those marked with $^\dagger$ or $^\ddagger$, the baryonic surface density refers to the baryonic surface density inside $|z|=1$~kpc and $|z|=1.1$~kpc, respectively. The baryonic surface density outside $|z|=1.1$~kpc is $3.3\MsunpcSq$ in \citet{McKee}, and $5.5\,\MsunpcSq$ at $|z|>1\,$kpc in \citet{Bovy:2012}. For our baryon model the baryonic surface density outside $|z|=1.0$~kpc is $8.2\,\MsunpcSq$, and outside $|z| = 1.1$~kpc is $7.3\,\MsunpcSq$. Legend from top to bottom refers to: \citet{Bienayme:2014}, \citet{Zhang:2013}, \citet{Xia:2016}, \citet{McKee}, \citet{Garbari:2012}, \citet{Bovy:2012}, \citet{Bovy:2013raa} and \citet{Smith:2012}}
\label{rhoDM_SigBa_tilt_fig}
\end{figure}

The motions of the tracer stars are dictated by the total distribution of mass in the galaxy, and are insensitive to the difference between baryons and DM. Thus there is a degeneracy between the baryons and the DM, and when comparing the local dark matter density results from different groups one should also take into account their values on the baryonic surface density. Figure~\ref{rhoDM_SigBa_tilt_fig} shows the result of an analysis including tilt of the $\alpha$-young data, but with a prior range on the total baryonic surface density that has been taken to be large enough to cover all ranges of dark matter densities and baryonic surface densities compatible with the $\alpha$-young tracer data (\ie allow a full error ellipse to form as opposed to having a hard cutoff). The red posterior ellipses of Fig.~\ref{rhoDM_SigBa_tilt_fig} clearly shows the tracer data degeneracy between baryonic surface density and dark matter density. Included in Fig.~\ref{rhoDM_SigBa_tilt_fig} are also published results from other authors, clearly showing that these results also follow a similar degeneracy between baryonic surface density and dark matter density. Hence, the apparent discrepancies between the results of different groups are not significant if this degeneracy between baryons and dark matter is taken into account.

\subsection{Taking into account the rotation curve term: $\mathcal{R}$} \label{rotation_curve_term_section}
So far in our analysis we have not taken into account the rotation curve term $\mathcal{R}$ from equation~(\ref{poisson_eq}). As in \cite{Silverwood:2015} the rotation curve term can from equation (\ref{poisson_eq}) be written as:
\begin{equation}
\mathcal{R}=\frac{1}{R}\frac{\partial V_c^2}{\partial R}=\frac{2V_c}{R}\frac{\partial V_c}{\partial R}=2(B^2-A^2), \label{Oort_const_eq}
\end{equation}
where $A$ and $B$ are the Oorts constants \citep[\eg][]{Binney2008}.

As in \cite{Silverwood:2015} we define an effective density $\rho_{\mathrm{eff}}(z)$, which incorporates the effective shift in density caused by the rotation curve term, \eg:
\begin{equation}
\frac{\partial^2 \Phi}{\partial z^2} = 4\pi G\rho_{\mathrm{eff}}(z), \label{eq:rho_eff_defn_context}
\end{equation}
where 
\begin{equation}
\rho_{\mathrm{eff}} = \rho(z) - \frac{1}{4\pi G R}\frac{\partial V_c^2}{\partial R}. \label{eq:rho_eff_defn} 
\end{equation}

From equations \ref{Oort_const_eq} and \ref{eq:rho_eff_defn} this effective density then relates to the true density $\rho(z)$ via  
\begin{equation}
\rho(z) = \rho_{\mathrm{eff}}(z) +\frac{B^2-A^2}{2\pi G}.
\end{equation}

\cite{Bovy_Prieto:2012} present measurements of the Milky Way's rotation curve using APOGEE data, and assuming the circular velocity to locally be a power law function of $R$ they find: 
\begin{align}
\frac{B^2-A^2}{2\pi G} &= 0.0002^{+0.0002}_{-0.0025}\, \Msunpc \\
&= 0.0076^{+0.0076}_{-0.095} \gevcm,  
\end{align}
and thus the true density is related to the effective density by:
\begin{align}
\rho &= \rho_{\mathrm{eff}} + 0.0002^{+0.0002}_{-0.0025} \Msunpc \\
& = \rho_{\mathrm{eff}} + 0.0076^{+0.0076}_{-0.095} \gevcm.
\end{align}
Hence the expected contribution from the rotation curve term is indeed quite small, an order of magnitude less than the statistical uncertainty for \eg the $\alpha$-young analysis. 

\section{Discussion} \label{discussion_section}

\subsection{The discrepancy between $\alpha$-young and $\alpha$-old populations} \label{sec:young_old_discrepancy}
As noted earlier in Section \ref{sec:results_full_analysis}, the $\rho_{\rm DM}$ posteriors derived from the $\alpha$-old and $\alpha$-young populations with tilt are discrepant to beyond the 95\% CR. The 95\% CR of the $\rho_{\rm DM}$ posterior from the combined analysis is within the 95\% CR of the $\alpha$-young population, but not that of the $\alpha$-old population.

In discussing these discrepancies we first look at the immediate data as it is before us in the plots, and how it leads to the divergent $\rhodm$ results. We will then discuss the root physical causes of the discrepancies such as incomplete modelling of the tilt term, a non flat rotation curve term, or the presence of disequilibria in the disc. 

\subsubsection{From $\nu$, $\sigma_z$, and $\sigma_{Rz}$ to discrepant $\rhodm$}
This discrepancy between the two populations was evident even in the simplified analysis presented in Section \ref{simplified_analytic_analysis_section}. While the $\alpha$-young population preferred ranges of $\rho_{\rm DM} = 0.02 \Msunpc = 0.76 \gevcm$ and $\Sigma_{\rm baryon}^\infty = 28 \MsunpcSq$ (or $\rho_{\rm DM} = 0.0145 \Msunpc = 0.55 \gevcm$ and $\Sigma_{\rm baryon}^\infty = 41 \MsunpcSq$ when neglecting the lower three data points), the $\alpha$-old population preferred a much lower DM density and a much higher baryon density, $\rho_{\rm DM} = 0.005 \Msunpc = 0.19\gevcm$ and $\Sigma_{\rm baryon}^\infty = 80 \MsunpcSq$ respectively. The $\alpha$-old baryon surface density in this case is almost double the number derived in Section \ref{sec:baryon_modelling}. 

In this simplified case all baryonic matter was concentrated in the midplane, and the $y$-intercept of the best-fit line was a measure of the baryonic surface density. Thus we can see that the source of this tension is that the $\sigma_z^2$ data for the $\alpha$-old population is flatter than expected, \ie~that one would expect the velocities of the $\alpha$-old tracer stars to be more strongly dependent on $z$. For the $\alpha$-old data the value of $\sigma_z$ rises by merely 12\% between the lowest and highest $z$ data point. For comparison, the value of $\sigma_z$ for the $\alpha$-young data rises with 23\% over a significantly shorter range in $z$.  
    
In the full \MultiNest analysis the $\alpha$-old population again prefers higher baryon surface densities, as seen in Fig. \ref{old_pop_fig}, where the posterior baryon distributions are straining against the upper edge of the prior (dashed green line). The posterior baryon distributions of the $\alpha$-young population, shown in Fig. \ref{young_pop_fig} are well centred in the prior range however, and while not as extreme as the $\alpha$-old only case, the posterior baryon surface density for the combined analysis is still very close to the upper edge of the prior, as seen in Fig. \ref{2pop_fig}. In contrast to the simplified analysis of Section \ref{simplified_analytic_analysis_section}, the $\rhodm$ posterior for the $\alpha$-old is roughly consistent with the $\alpha$-young data when neglecting tilt, but is \textit{higher} than the $\alpha$-young result when including tilt. We will now take a closer look at the tilt term and how it could impact this discrepancy. 

\subsubsection{The role of tilt in the $\alpha$-young -- $\alpha$-old discrepancy}\label{young_old_discrepancy_tilt}
In \cite{Silverwood:2015} we illustrated the necessity of including the tilt term in determinations of $\rhodm$, especially for populations with larger scale heights. One potential addendum to that conclusion is that while including the tilt term is necessary, we do not yet have the necessary data to properly model and calculate it. 

Here we model the tilt term with three parameters, $A$, $n$, and $k$, as introduced in Section \ref{tilt_term_section} and specifically equation (\ref{tilt_eq}). When analyzing the posterior distributions of the tilt terms $A$ and $n$ we find that their distributions are nicely peaked and confined well within their prior ranges. This is because the $A$ and $n$ parameters are used to model $\sigma_{Rz}(z)$, which is fit to data from B15. Hence these parameters are constrained by data, and as anticipated in Section \ref{n_prior}, their prior ranges are generous enough to not impact our results. 

In contrast, the tilt parameter $k$, which encodes information about the radial variation of $\nu$ and $\sigma_{Rz}$, is not constrained by the data as B15 has $\nu$ and $\sigma_{Rz}$ data only at the Solar radius. Thus we have to enforce a prior on this parameter, as discussed in Section \ref{K_prior}. The $k$ tilt parameter enters through $K\equiv (1/R_\odot-k)$, which in turn is a proportionality constant in calculating $\sigma_{z,\mathcal{T}}^2$ (see equation \ref{sigz2_tilt}), making $K$ a gauge on the size and sign of the tilt term. For the analysis of the $\alpha$-young data only, the posterior $k$ distribution is more or less flat within its prior ranges. On the other hand the $\alpha$-old data have some tension in them which appears to drive the $k$ posterior distribution towards one of the edges of the prior. This calls into question the reliability of the tilt model and prior for the $\alpha$-old population, and hence also the reliability of the resulting $\rhodm$ posterior.  

To adequately address this issue we need radial information on $\nu$ and $\sigma_{Rz}$ from the same stellar data set as used to derive the other quantities used in the analysis. With this we can properly model the tilt term and fit it to real data in a consistent manner, as opposed to imposing a prior from a potentially ill-suited external source. The upcoming second data release of Gaia holds the promise of providing such information.

\subsubsection{The role of disequilibria}\label{sec:disequilibria}
Determinations of $\rhodm$ made so far, including this analysis, have assumed that the disc is in dynamical equilibrium, and thus all time derivatives disappear. However there is evidence that there are time-dependent disequilibria which break this assumption \citep[see \eg][]{2012ApJ...750L..41W, 2013MNRAS.429..159G, 2013MNRAS.436..101W, 2013ApJ...777L...5C, 2014MNRAS.440.1971W, 2017A&A...602L..13A}. At the very least the Milky Way has spiral arms, which would break not only dynamical equilibrium but also axisymmetry at the Solar position. Also note that both our method and the derivation of the input data from B15 assume symmetry about $z=0$, an assumption broken by many examples of disequilibria, adding another avenue for disequilibria to enter and impact our analysis.  

Additionally, these disequilibria could have different impacts on different stellar populations. For instance stars with higher scale heights have longer crossing times and so longer equilibration times \citep{Binney2008}, and thus such populations could still be exhibiting the effects of a long past perturbative event, while those with lower scale heights and shorter crossing/equilibration times had returned to equilibrium.

Recently \cite{2017MNRAS.464.3775B} investigated the impact of such disequilibria on determinations of the local DM density and demonstrated that if present, neglecting them could results in a 25\% error on the recovered $\rhodm$ value. Interestingly they also concluded that a sign of such disequilibria could be different stellar populations yielding divergent $\rhodm$ figures, which has occured with our analysis. They also note however that this effect would be difficult to disentangle from other effects such as the tilt term.

\subsubsection{The role of the rotation curve term}
Finally, our analysis also assumes that the rotation curve is locally flat, and thus the rotation curve term in Eq. \ref{poisson_eq} is zero. If this assumption does not hold then this can manifest as a systematic shift in the local DM density. In Section \ref{rotation_curve_term_section} we estimate that the size of this shift is small, using results from the APOGEE survey \citep{Bovy_Prieto:2012}. However this estimation (from which we justified our assumption of the locally flat rotation curve) is made based on observations from within a few hundred parsecs of the mid plane. At larger heights above the disc the rotation curve may become increasingly non-flat, contributing to the discrepancy between between the $\alpha$-old and $\alpha$-young populations. This should be investigated with upcoming Gaia data.

\subsubsection{The role of observational uncertainties}
The $\alpha$-old population reaches to a higher $z$ above the disc plane (and so further from the Sun), and thus is more susceptible to distance errors and contamination from halo stars. While B15 took great care to exclude interlopers it is possible that errors in this process have biased the high-$z$ results and contributed to the discrepancy. 

Additionally there is the possibility of problems with the metallicity cuts made to extract the $\alpha$-young and $\alpha$-old populations. Our results concur with \cite{2015A&A...579A.123H}, who previously noted the inconsistency of the two populations, but was unable to reconcile the two with the inclusion of tilt. Instead they argued that the stars in each sample were insufficiently homogenous for a simple kinematic analysis to be used. Populations with different velocity dispersions could be contaminating the samples, increasing or decreasing the slope of the $\sigma_z$ data points. With Gaia we will have a much larger number of stars at our disposal, allowing for smaller metallicity bins and a reduction of any possible cross contaminations between populations.

\subsection{Comparison of $\rhodm$ from $\alpha$-young and $\alpha$-old results}
From the previous section investigating the discrepancy between the $\alpha$-young and $\alpha$-old results we can see that the problems identified weigh most heavily upon the $\alpha$-old population. The tilt term has a greater impact on the $\alpha$-old population, as witnessed by the difference in $\rhodm$ results when neglecting or including tilt (see Table~\ref{tab:results}). Thus it also experiences a greater impact from our assumptions and priors on the tilt term. In contrast the $\alpha$-young result changes very little when the tilt term is included or neglected, and is less dependent on the tilt model and its assumptions. Furthermore while the $\alpha$-young $k$ posterior is largely flat within its prior boundaries, the $\alpha$-old $k$ posterior is heavily biased towards the edge of its prior, hinting at an underlying tension in the model. 

Populations with lower scale heights have shorter equilibration times, and so if there are disequilibria present in the disc from past perturbative events it is likely they will have less impact on the $\alpha$-young result than the $\alpha$-old result. 

The assumption of a flat rotation curve has been made based on local populations, and thus lower scale heights. Thus this assumption is more valid for the $\alpha$-young population than for the $\alpha$-old population which has a higher scale height. 

Finally, the problems of halo contamination and distance errors are more acute for the $\alpha$-old population compared to the $\alpha$-young population.

Thus with these points in mind we can conclude that the most trustworthy results are those derived from the $\alpha$-young population. Given that the posteriors for the tilt terms $A$ and $n$ are driven by the data, and the posterior for $k$ is flat within its data-derived prior, we also conclude that the the best $\rhodm$ result is that derived from the $\alpha$-young with tilt, namely $\rhodm = 0.46^{+0.07}_{-0.09}\, \gevcm = 0.012^{+0.001}_{-0.002} \Msunpc$ (68\% uncertainty limits). 

\section{Conclusions}
In this work we have applied the integrated Jeans equation method first presented in \cite{Silverwood:2015} to SDSS-SEGUE G-dwarf data from \cite{Budenbender} in order to determine the local DM density. This quantity is of vital importance to interpreting results, be they positive or negative, from a wide range of DM detection experiments.

The analysis featured an multi-component baryon model, given in Section \ref{sec:baryon_modelling}, and included detailed modelling of the tilt term describing the radial-vertical coupling of the stellar motions, discussed in Section \ref{tilt_term_section}. From the SDSS-SEGUE data \cite{Budenbender} extracted two populations based on metallicity, named $\alpha$-young and $\alpha$-old. The former has a lower scale height and can be considered to populate the canonical `thin disc', while the latter has a higher scale height and so populates the canonical `thick disc'. These two populations were analysed separately, with and without the modelling of tilt, and also together with tilt modelling, sharing a common DM and baryon mass models but separate tracer density, vertical velocity dispersions and tilt models. These analyses, summarised in Table~\ref{tab:results}, showed a discrepancy in derived $\rhodm$ between the two populations, the cause of which was discussed in Section \ref{sec:young_old_discrepancy}. The impacts of all the underlying causes considered fall most heavily upon the $\alpha$-old population, while the $\alpha$-young results remain robust. Thus we conclude from the analysis of the $\alpha$-young data with tilt that the local DM density is $\rhodm = 0.46^{+0.07}_{-0.09}\, \gevcm = 0.012^{+0.001}_{-0.002} \Msunpc$ (68\% uncertainty limits).

This result is built upon the assumption of dynamical equilibrium. However there is evidence of disequilibria present in the disc, as discussed in Section \ref{sec:disequilibria}. Theoretical estimates from \cite{2017MNRAS.464.3775B} suggest such disequilibria, if ignored, could bias the local DM density estimates by 25\%. In future work we will investigate disequilibria further and incorporate them into our determinations of the local DM density.  

Along with the rest of the astronomical and astrometric community we await the release of Gaia Data Release 2, scheduled for April 2018. The priorities for this upcoming data set in regards to the local DM density will be to derive $\nu$, $\sigma_z$, and $\sigma_{Rz}$ data not just at the solar radius $R_\odot$ but also at radii inside and outside this radius. This will allow us to self-consistently model and fit the tilt term to data, and also potentially to apply our method to other regions of the galaxy and determine the DM density there. A measurement of the slope of the rotation curve above and below the disc will also be important to determine the rotation curve term in the Poisson equation (Eq. \ref{poisson_eq}). Also, given the degeneracy between the baryon surface density and $\rhodm$ illustrated in Fig. \ref{rhoDM_SigBa_tilt_fig} and discussed in Section \ref{sec:dm_baryon_degeneracy}, it will be necessary to supplement Gaia data with further observations aimed at mapping the local baryon distribution more accurately. We note that a great deal of the disagreement between previous $\rhodm$ measurements can be attributed to differences in their respective baryon surface densities. This is illustrated in Figure \ref{rhoDM_SigBa_tilt_fig}, where the many $\rhodm$ determinations lie along the same $\rhodm -\, \Sigma_{\rm baryon}$ band defined by our posterior ellipses (red and crimson). While Gaia data will no doubt improve the mass distribution of visible stars, a sizeable fraction of the baryon density uncertainty arises from gas, a component beyond the reach of Gaia.

\section*{Acknowledgments} 
We would like to thank Katherine Freese and Glenn van de Ven for their support, advice, and discussions throughout this project. S.S. acknowledges support from the Swedish Research Council (VR, Contract No. 350-2012-268 and No. 638-2013-8993). H.S. acknowledges support from the MINECO (Spanish Ministry of Economy) - FEDER through grant MDM-2014-0369 of ICCUB (Unidad de Excelencia `Maria de Maeztu'). J.I.R. would like to acknowledge support from STFC consolidated grant ST/M000990/1 and the MERAC foundation. G.B. acknowledges support from the European Research Council through the ERC starting grant \textit{WIMPs Kairos}.

\bibliographystyle{mnras}
\bibliography{LocalDM_Bibliography}


\appendix

\section{Prior range on tilt parameters \lowercase{$k_0$} and \lowercase{$k_1$}}  \label{k1_prior_appendix}
\begin{figure}
\includegraphics[scale=0.42, trim={0 0 0.3cm 0},clip]{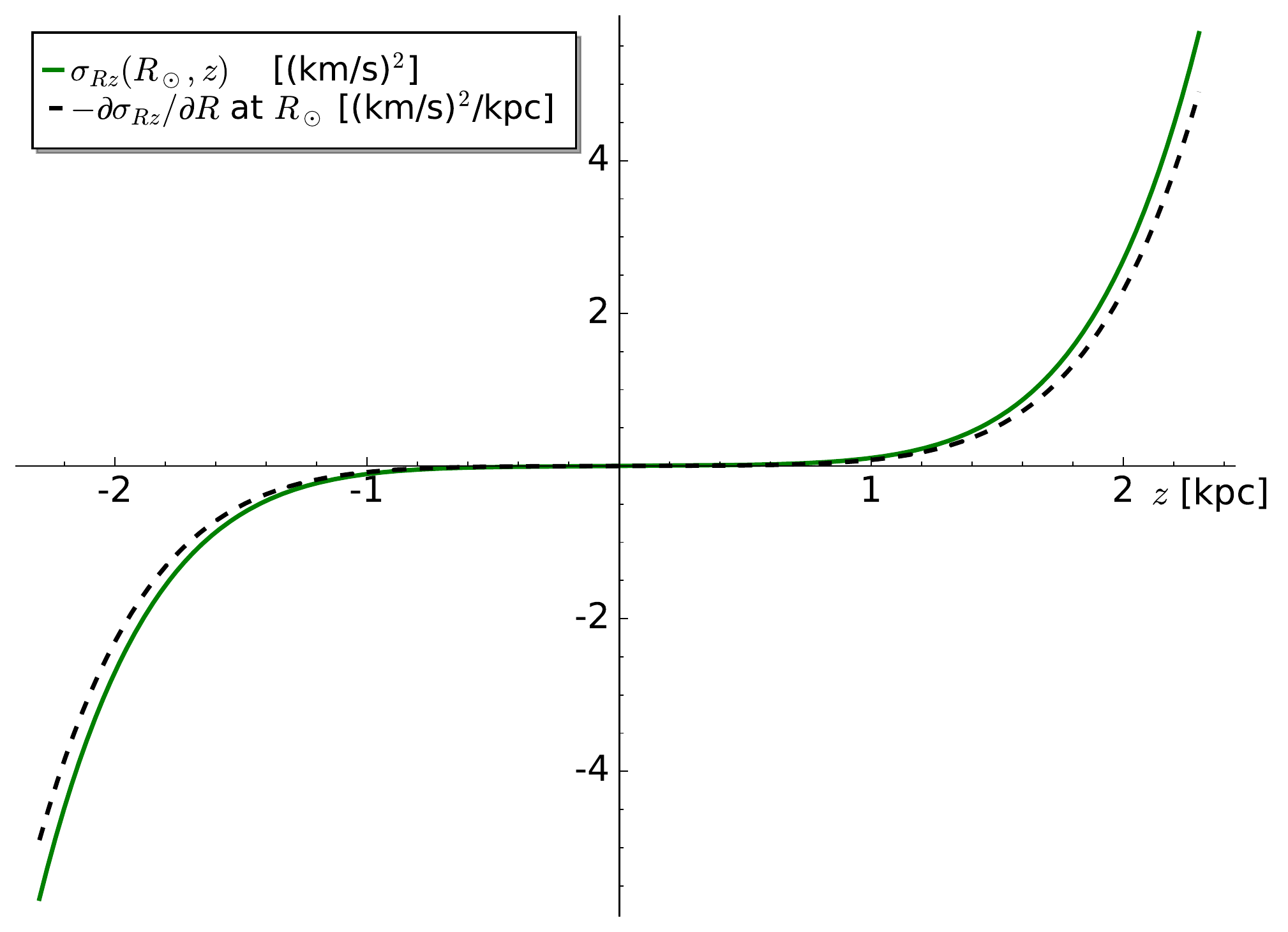} 
\caption{$\sigma_{Rz}$ (green solid line) and $-\partial\sigma_{Rz}/\partial R$ (black dashed line) as functions of $z$ at $R=R_\odot$. For hot dwarfs from \protect\cite{Binney:2014}.}
\label{RAVE_sigma_-dsigmadR_fig}
\end{figure}

As we don't have data on how $\sigma_{Rz}$ behaves with $R$ for our tracer stars, one way to see if our assumptions are reasonable is by looking at the result from other stellar catalogs. In \cite{Binney:2014} they parametrize $\sigma_1$, $\sigma_2$ and $\theta$ as functions of $R$ and $z$ and then fit to RAVE data. Here $\sigma_1$ and $\sigma_2$ are the semi-major and semi-minor axes, respectively, of the velocity ellipse in the $R,z$ -plane; $\theta$ denotes the angle between semi-major axis and the galactic plane. For more details, see \cite{Binney:2014}. From the result of \cite{Binney:2014} we can derive $\sigma_{Rz}$ through
\begin{equation}
\sigma_{Rz}=(\sigma_1^2-\sigma_3^2)\sin(\theta)\cos(\theta). \label{appendix_sigmaRz_equation}
\end{equation}
For the hot dwarf population of \cite{Binney:2014} the resulting $\sigma_{Rz}(R_\odot,z)$ and $-\frac{\partial \sigma_{Rz}}{\partial R}(R_\odot,z)$ are shown in Fig.~\ref{RAVE_sigma_-dsigmadR_fig}. From our $k_1$ modelling in Section \ref{tilt_term_section} we have that
\begin{equation}
k_1=\frac{-1}{\sigma_{Rz}(R_\odot,z)} \frac{\partial \sigma_{Rz}}{\partial R}(R_\odot,z). \label{appendix_k1_equation}
\end{equation}
Hence Fig.~\ref{RAVE_sigma_-dsigmadR_fig} immediately shows that for this data set we have $0\leq k_1\leq 1$ kpc$^{-1}$, \ie within our $k_1$ prior range as stated in Section \ref{K_prior}: $-1\leq k_1\leq 1$ kpc$^{-1}$.

From Fig.~\ref{RAVE_sigma_-dsigmadR_fig} we see that the shape of $\sigma_{Rz}$ for the hot dwarfs of \cite{Binney:2014} does not match the $\sigma_{Rz}$ shape of the tracer stars from B15 that we use. Note however that both $\sigma_{Rz}$ functions are well fit by functions of the form $Az^n$. From Fig.~\ref{RAVE_sigma_-dsigmadR_fig} we see that $\sigma_{Rz}$ carries the same sign as $z$, as predicted in Section \ref{tilt_term_section}, and that the shapes of the green solid and black dashed lines are very similar. The latter observation supports the assumption in Section \ref{tilt_term_section} that $k_1$, as given in equation (\ref{appendix_k1_equation}), is reasonably constant for all relevant $z$ values. \footnote{The $z$ region of practical interest to us, \ie the region of $\sigma_{Rz}$ data, is: $z=$515 pc to 1247 pc for the $\alpha$-young population, and $z=$634 pc to 2266 pc for the $\alpha$-old population. Over these regions the value of $k_1$ increases with $z$ by 42\% for the $\alpha$-young data range, and by 37\% for the $\alpha$-old data range.}

Another source of information on the nature of the $\sigma_{Rz}$ dependance on $R$ is by looking at evolved N-body mock data. One such mock data set is described in \cite{Garbari:2011}, whose evolution takes into account the formation of a bar and spiral arms, and their effects on the distribution functions. We have extracted $\sigma_{Rz}$ as a function of $R$ for different values of $z$ and azimuthal angle $\varphi$ for the evolved mocks from \cite{Garbari:2011}. From the extracted $\sigma_{Rz}$ plots (not shown) we find that $\sigma_{Rz}$ from \cite{Garbari:2011} is a much more flat function of $R$, \ie corresponds to a much smaller value on $k_1$, than was the case for $\sigma_{Rz}$ from \cite{Binney:2014}. Due to statistical noise it is not possible to reliably extract the small value on the $\sigma_{Rz}$ slope, and hence the value on $k_1$, from the \cite{Garbari:2011} mock data; we however conclude that the mock data does support our notion that $k_1$ is expected to have a sufficiently small value, which is in agreement also with the \cite{Binney:2014} result.

\bsp	
\label{lastpage}
\end{document}